\documentclass[aps,prl,twocolumn,preprintnumbers,amsmath,amssymb,superscriptaddress
]{revtex4-2}
\usepackage{amsmath,graphicx}
\usepackage[utf8]{inputenc}
\usepackage[T1]{fontenc}
\usepackage{xcolor}
\usepackage{textcomp}
\usepackage{bm}
\usepackage{siunitx}
\usepackage{physics}
\usepackage{amsmath}
\usepackage{tikz}
\usepackage{mathdots}
\usepackage{yhmath}
\usepackage{cancel}
\usepackage{color}
\usepackage{array}
\usepackage{multirow}
\usepackage{amssymb}
\usepackage{gensymb}
\usepackage{tabularx}
\usepackage{extarrows}
\usepackage{booktabs}
\usetikzlibrary{fadings}
\usetikzlibrary{patterns}
\usetikzlibrary{shadows.blur}
\usetikzlibrary{shapes}
\usepackage{makecell}

\usepackage{color}
\definecolor{LinkColor}{rgb}{0.75,0.0,0.2}

\usepackage{hyperref}
\hypersetup{
	pdfauthor={good guys},
	pdftitle={good title},
	colorlinks=true,
	citecolor=LinkColor,
	linkcolor=LinkColor,
	urlcolor=LinkColor,
}

\usepackage{listings}
\definecolor{lightgray}{gray}{1}

\begin{document}

\title{Theory of Topological Superconductivity and Antiferromagnetic Correlated Insulators in Twisted Bilayer WSe${}_2$}
\author{Chuyi Tuo}
\thanks{These two authors contribute equally in this work}
\affiliation{Institute for Advanced Study, Tsinghua University, Beijing 100084, China}
\author{Ming-Rui Li}
\thanks{These two authors contribute equally in this work}
\affiliation{Institute for Advanced Study, Tsinghua University, Beijing 100084, China}
\affiliation{Department of Physics, Princeton University, Princeton, New Jersey 08544, USA}
\author{Zhengzhi Wu}
\affiliation{Institute for Advanced Study, Tsinghua University, Beijing 100084, China}
\affiliation{Rudolf Peierls Centre for Theoretical Physics, Parks Road, Oxford, OX1 3PU, UK}
\author{Wen Sun}
\affiliation{Institute for Advanced Study, Tsinghua University, Beijing 100084, China}
\author{Hong Yao}
\email{yaohong@tsinghua.edu.cn}
\affiliation{Institute for Advanced Study, Tsinghua University, Beijing 100084, China}
\date{\today}

\maketitle

\section{Abstract}
Since the very recent discovery of unconventional superconductivity in twisted WSe${}_{2}$ homobilayers at filling $\nu=-1$, considerable interest has arisen in revealing its mechanism. In this paper, we developed a three-band tight-binding model with non-trivial band topology by direct Wannierization of the low-energy continuum model. Incorporating both onsite Hubbard repulsion and
next-nearest-neighbor attraction,  
we then performed a mean-field analysis of the microscopic model and obtained a phase diagram qualitatively consistent with the experiment results. For zero or weak displacement field, the ground state is a
Chern number $C=\pm 2$ topological superconductor in the Altland-Zirnbauer A-class (breaking time-reversal but preserving total $S_z$ symmetry) with inter-valley pairing 
dominant in $d_{xy}\pm id_{x^2-y^2}$-wave (mixing with a subdominant $p_x\mp i p_y$-wave) component. 
For a relatively strong displacement field, the ground state is a correlated insulator with the $120^\circ$ antiferromagnetic order. Our results provide new insights into the nature of the twisted WSe${}_{2}$ systems and suggest the need for further theoretical and experimental explorations.

\section{Introduction}
In recent years, twisted van der Waals moiré superlattices~\cite{andrei2020graphene,andrei2021marvels,kennes2021moire,castellanos2022van,mak2022semiconductor} have garnered significant interest following the groundbreaking discovery of unconventional superconductivity (SC)~\cite{cao2018unconventional,lu2019superconductors,yankowitz2019tuning,oh2021evidence} in magic-angle twisted bilayer graphene~\cite{bistritzer2011moire}, which inspired numerous theoretical investigations on its microscopic origin~\cite{PhysRevLett.121.257001,PhysRevLett.122.257002,PhysRevB.99.165112,PhysRevLett.127.247703,PhysRevB.110.045133,wang2024molecular,wang2024electron,PhysRevB.98.081102,PhysRevB.105.L100503,shavit2021theory,PhysRevX.8.041041,PhysRevX.9.041010,PhysRevB.106.235112,christos2023nodal,PhysRevB.107.L060503,PhysRevLett.129.047601,dodaro2018phases,you2019superconductivity,khalaf2021charged,lothman2022nematic}. 
Beyond graphene-based systems~\cite{cao2018correlated,park2021tunable,hao2021electric,cao2021pauli,kim2022evidence,zhou2022isospin,zhang2023enhanced,arora2020superconductivity,holleis2024nematicityorbitaldepairingsuperconducting,li2024tunable,su2023superconductivity,zhou2021superconductivity,han2024signatures,choi2024electric}, twisted transition metal dichalcogenides (TMDs)~\cite{PhysRevLett.121.026402,PhysRevLett.122.086402} are regarded as a promising alternative platform to investigate many-body physics.  Due to their high tunability, twisted TMDs can host a wide variety of exotic phases, including correlated insulators~\cite{tang2020simulation,regan2020mott,wang2020correlated,xu2020correlated,li2021continuous,ghiotto2021quantum,xu2022tunable}, integer and fractional quantum anomalous Hall states~\cite{li2021quantum,cai2023signatures,zeng2023thermodynamic,park2023observation,xu2023observation}, and also integer and fractional quantum spin Hall states~\cite{kang2024evidence}. However, the experimental realization of SC in twisted TMD systems has remained elusive~\cite{wang2020correlated}, though many theoretical works have suggested that SC should develop in such systems~\cite{PhysRevResearch.4.043048,PhysRevResearch.5.L012034,PhysRevB.108.155111,PhysRevB.106.235135,PhysRevB.108.L201110,PhysRevB.108.064506,akbar2024topological,PhysRevLett.130.126001,PhysRevLett.131.016001,schrade2024nematic,crepel2023topological}, typically through doping the correlated insulators. Very recently, two independent studies have reported the discovery of robust SC phases in twisted WSe${}_{2}$ (tWSe${}_{2}$) homobilayers~\cite{xia2024unconventional,guo2024superconductivitytwistedbilayerwse2}.  
The phenomenology of these SC phases differ significantly from that would arise in doped Mott insulators, which calls for further theoretical investigations, particularly regarding the possible pairing mechanism  
and topological properties of SC, as well as the nature of adjacent correlated insulators.

Previous theoretical studies~\cite{wang2020correlated,PhysRevResearch.2.033087,PhysRevB.104.075150,10.1063/5.0077901,PhysRevResearch.4.043048,PhysRevResearch.5.L012034,PhysRevLett.130.126001,PhysRevB.106.235135,PhysRevB.108.L201110,PhysRevB.108.155111,PhysRevB.108.064506,akbar2024topological,pan2020quantum} of tWSe$_{2}$ homobilayers are largely based on the tight-binding description of the single-band moiré Hubbard model with spin-dependent hopping phase tuned by the displacement field.  
By formulating the problem in real space, this approach captures the locality of screened interactions and facilitates a clearer real-space insights essential for understanding the correlated physics. 
However, such a simple single-band model cannot capture the potential nontrivial band topology~\cite{PhysRevLett.122.086402,PhysRevB.105.L081108,zhang2024polarization,foutty2024mapping,devakul2021magic} and only applies to limited twist angles~\cite{wang2020correlated}. 
Thus, a direct Wannierization~\cite{devakul2021magic,PhysRevX.13.041026,xu2024maximally,crepel2024bridging,PhysRevX.8.031087,PhysRevX.8.031088,PhysRevB.99.195455} of the standard continuum model~\cite{PhysRevLett.122.086402} with a proper set of low-energy bands appears to be a more suitable approach to start with.

In this paper, we focus mainly on the experimental results in Ref.~\cite{xia2024unconventional}, where the SC phase is observed in a $3.65^\circ$ tWSe${}_2$ device at integer filling factor $\nu=-1$ under a small displacement field, along with a correlated insulator phase in a larger displacement field. Using the continuum model parameters provided in Ref.~\cite{devakul2021magic}, we first construct a three-band tight-binding model for $3.65^\circ$ tWSe${}_2$ through direct Wannierization~\cite{devakul2021magic,PhysRevX.13.041026,xu2024maximally,crepel2024bridging,PhysRevX.8.031087,PhysRevX.8.031088,PhysRevB.99.195455}. In addition to the onsite Hubbard repulsion, our model  
includes the next-nearest-neighbor (NNN) attraction, which can arise effectively through electron-phonon coupling 
\cite{bardeen1957theory,mcmillan1968transition,
PhysRevLett.121.257001,PhysRevLett.122.257002,PhysRevB.99.165112,PhysRevLett.127.247703,PhysRevB.110.045133,wang2024molecular,wang2024electron,PhysRevX.8.041041,PhysRevX.9.041010,PhysRevB.106.235112,christos2023nodal,PhysRevB.107.L060503,PhysRevB.98.081102,PhysRevB.105.L100503,shavit2021theory} or 
purely electronic mechanisms~\cite{cea2021coulomb,PhysRevLett.127.247001,PhysRevB.97.235453,PhysRevB.99.134515,chatterjee2022inter,kohn1965new,monthoux1991toward,scalapino1995case}.  
By keeping these key interaction terms, our model can capture the essential qualitative physics of the experiment~\cite{xia2024unconventional} and offer valuable real-space insights into the twisted WSe$_2$ system. We then perform a thorough mean-field analysis to the interacting model and obtain the phase diagram at filling $\nu=-1$ under realistic interaction strengths, concluding that the SC phase is consistent with inter-valley pairing with mixed $d_{xy}\pm id_{x^2-y^2}$ and $p_x\mp ip_y$-wave symmetry, 
featuring nontrivial topology under Altland-Zirnbauer A-class~\cite{PhysRevB.55.1142,schnyder2008classification,chiu2016classification} with Chern number $C=\pm 2$, and the correlated insulator phase has $120^\circ$ antiferromagnetic (AFM) order. Additionally, the mixed symmetry character of the SC phase near zero displacement field predicted by our model has experiment observable distinctions from that of moiré Hubbard model~\cite{wang2020correlated,PhysRevResearch.2.033087,PhysRevB.104.075150,10.1063/5.0077901,PhysRevResearch.4.043048,PhysRevResearch.5.L012034,PhysRevLett.130.126001,PhysRevB.106.235135,PhysRevB.108.L201110,PhysRevB.108.155111,PhysRevB.108.064506,akbar2024topological,pan2020quantum, PhysRevB.88.041103,PhysRevB.100.060506} due to the absence of emergent spin-valley SU(2) symmetry~\cite{PhysRevResearch.2.033087,PhysRevB.104.075150}, providing new insights into the nature of the moiré TMD systems.

\section{Results}
\textit{Continuum model and symmetry analysis.}---We begin with the standard low-energy continuum model description of tWSe${}_2$~\cite{PhysRevLett.122.086402}. The non-interacting continuum Hamiltonian for the $K$ valley is given by: 
\begin{equation}
H_K(\bm{r})=\left(\begin{array}{cc}
-\frac{\hbar^2\left(\bm{k}-\boldsymbol{\kappa}_{+}\right)^2}{2 m^*}+\Delta_+(\bm{r}) & \Delta_{\text{T}}(\bm{r}) \\
\Delta_{\text{T}}^{\dagger}(\bm{r}) & -\frac{\hbar^2\left(\bm{k}-\boldsymbol{\kappa}_{-}\right)^2}{2 m^*}+\Delta_-(\bm{r})
\end{array}\right)
\end{equation}
where $m^*$ is the effective mass of valence band edge, $\bm{\kappa}_\pm$ are located at corners of the mini Brillouin zone as shown in Fig.~\ref{fig:fitband}(a).
In the lowest-order harmonic approximation, we only keep terms with moiré wave vectors $\bm{g}_j$ which are obtained by rotation of $\bm{g}_1 = (\frac{4\pi}{\sqrt{3}a_M}, 0)$ by $(j-1)\pi/3$ in moiré potential $\Delta_\pm(\bm{r})$ and interlayer tunneling $\Delta_{\text{T}}(\bm{r})$:
\begin{align}\label{moire_potential_and_tunneling}
\Delta_{\pm}(\bm{r}) &= \pm \frac{V_z}{2} + 2 V \sum_{j=1,3,5} \cos \left(\bm{g}_j \cdot \bm{r} \pm \psi\right)\\
\Delta_{\text{T}}(\bm{r}) &= w\left(1+e^{-i \bm{g}_2 \cdot \bm{r}}+e^{-i \bm{g}_3 \cdot \bm{r}}\right)
\end{align}
and the Hamiltonian for $-K$ valley is related by time-reversal symmetry $\mathcal{T}$. Throughout this paper, we adopt the experimental relevant twist angle $\theta=3.65^\circ$~\cite{xia2024unconventional}, continuum model parameters derived by large scale ab initio simulations ($V$, $\psi$, $w$) = ($9$ meV, $128^\circ$, $18$ meV)~\cite{devakul2021magic} and the effective mass $m^*=0.45m_e$~\cite{PhysRevLett.116.086601}. The moiré lattice constant is given by $a_M \approx a_0/\theta$ with $a_0=3.317$\AA~\cite{mounet2018two}.

In the absence of displacement field $V_z=0$, tWSe${}_2$ system has $D_3$ point group symmetry generated by threefold rotation $C_{3z}$ and twofold rotation $C_{2y}$. Additionally, due to the lowest order harmonic approximation we adopted, the continuum model has additional pseudo-inversion symmetry $\mathcal{I}$ with $\sigma_x H_K(\bm{r})\sigma_x = H_K(-\bm{r})$, enlarging the point group symmetry to $D_{3d}$. When a finite displacement field $V_z \neq 0$ is applied, the point group symmetry is reduced from $D_{3d}$ to $C_{3v}$, breaking all symmetries that interchange the two layers. Apart from point group symmetries, the system also exhibits U(1) spin-valley symmetry and time-reversal symmetry $\mathcal{T}$ for both $V_z=0$ and $V_z\neq 0$. These symmetry considerations are crucial for the subsequent construction of tight binding model and the classification of SC pairing symmetries.

\textit{Wannier functions and tight-binding model.}---To study the low-energy physics, we first focus on the top moiré valence bands of the tWSe${}_2$ continuum model at $V_z=0$.  
The Chern numbers for each valley share the same sign up to top five moiré bands, precluding any low-energy real-space description with correct band topology due to Wannier obstructions. Therefore, we focus on accurately reproducing the Chern numbers of the top two moiré bands, and construct a minimal three-band tight-binding model with Chern numbers ($1,1,-2$) for $K$ valley.
Following Refs.~\cite{devakul2021magic, PhysRevX.13.041026,xu2024maximally,crepel2024bridging}, we construct sufficiently localized Wannier functions using their layer polarization properties and the location of Wannier centers. 
The resulting Wannier functions are shown in Fig.~\ref{fig:fitband}(b), where the A and B orbitals are centered at XM and MX regions $(\pm a_M/\sqrt{3},0)$ with opposite layer polarization, and the C orbital is centered at MM region $(0,0)$ with layer hybridization. Detailed symmetry analysis~\cite{xu2024maximally} reveals that the (A, B, C) orbitals for the $K$ valley have $C_{3z}$ eigenvalues ($e^{-i2\pi/3}$, $e^{-i2\pi/3}$, $1$), 
and $C_{2y}\mathcal{T}$ or $\mathcal{I}$ symmetries interchange A and B orbitals but leave C orbital invariant. Using these Wannier functions, we can construct the following tight-binding model:
\begin{equation}
    H_0 = \sum_{i\alpha j\beta\sigma}t_{i\alpha j\beta \sigma} c^\dagger_{i\alpha\sigma} c_{j\beta\sigma}\label{equ:tightHam}
\end{equation}
where $i,j$ are unit cell indexes, $\alpha,\beta$ are sublattice indexes (A, B, C), and $\sigma$ is spin-valley index $\uparrow$ (or $K$), $\downarrow$ (or $-K$). By keeping up to fifth-nearest-neighbor hopping parameters (see Sec.~I of the Supplementary Information~\cite{supp}), Fig.~\ref{fig:fitband}(c) illustrates the close matching between the band structure of the tight-binding model and the continuum model, especially for the top moiré valence band.

We then consider the case of $V_z\neq 0$. Instead of repeating the above procedure for each $V_z$ separately, we utilize the layer polarization properties of Wannier functions, modeling the effect of displacement field as:
\begin{equation}\label{hd}
    H_D = \frac{\mathcal{V}_z}{2}\sum_{i}(n_{iA} - n_{iB})
\end{equation}
where $\mathcal{V}_z$ is the energy difference between A and B sublattices induced by the displacement field, and $n_{i\alpha}=\sum_{\sigma} c_{i\alpha\sigma}^\dagger c_{i\alpha\sigma}$ is the density operator. Since the energy expectation value of displacement field on the C sublattice can be chosen as $0$, we neglect terms involving it in $H_D$ (see Sec.~II of the Supplementary Information~\cite{supp} for a more detailed discussion of the displacement field effects). The Fermi surfaces of filling factor $\nu=-1$ with $\mathcal{V}_z=0$, $5$, $15$ and $25$ meV are shown in Fig.~\ref{fig:fermisurface}(a), where the Fermi surface of spin up and down are split by the displacement field $\mathcal{V}_z$, but related by the time-reversal symmetry $\mathcal{T}$. Fig.~\ref{fig:fermisurface}(b) also illustrates the Fermi surface density of states (DOS) as a function of 
$\mathcal{V}_z$.

To capture the many-body physics, we consider leading interactions in tWSe${}_2$. The dominant interaction 
should be the onsite Hubbard repulsion:
\begin{equation}
    H_U = \sum_{i\alpha} U_{\alpha} n_{i\alpha\uparrow} n_{i\alpha\downarrow} 
\end{equation}
where $\alpha=A,B,C$ labels the sublattice, and $n_{i\alpha\sigma}=c_{i\alpha\sigma}^\dagger c_{i\alpha\sigma}$. 
The positive Hubbard interaction can typically lead to magnetic ordering,  
providing a promising explanation of the correlated insulator phase. Longer-range Coulomb repulsions are neglected mainly because their magnitudes are much smaller (see Sec.~IV of the Supplementary Information~\cite{supp}).

Due to the time-reversal symmetry $\mathcal{T}$ of the systems, the Fermi surface features Cooper instability; namely, SC instabilities can occur even with infinitesimal attractions. 
Considering that the Fermi surface shown in Fig.~\ref{fig:fermisurface}(a) favor pairing within the same sublattices and has almost no C-orbital component, a natural choice is to consider the attractions 
on NNN sites of the same A or B sublattices:
\begin{equation}
    H_{V_2} = -V_2 \sum_i \sum_{\alpha\in \{A,B\}} \sum_{\delta\in \text{NNN}} n_{i+\delta \alpha} n_{i\alpha} 
\end{equation}
where $V_2$ is the strength of NNN attraction, $\alpha$ is sublattice index of A and B, and $\delta$ represents one of the three NNN bond directions $120^\circ$ apart.   
Besides, we have also considered the effects of nearest-neighbor (NN) attraction between A and B sublattices in Sec.~VI of the Supplementary Information~\cite{supp}, where we have shown that it is less dominant. 
The relatively local NNN (or NN) attraction can be understood by a random phase approximation (RPA) analysis (see Sec.~V of the Supplementary Information~\cite{supp}). 
Therefore, the interacting model which we should consider to describe the main physics of tWSe$_2$ is denoted as $H = H_0 + H_D + H_U + H_{V_2} $.

\textit{Mean-field analysis of SC and AFM}--- 
We start with the mean-field analysis of SC instabilities by decoupling the NNN attraction $H_{V_2}$ in the SC channel as:
\begin{eqnarray}
      &&H_{V_2}
        \approx -V_2\sum_{i \sigma \sigma^\prime} \sum_{\alpha\in \{A,B\}} \sum_{\delta\in NNN} \bigg( 
      \tilde{\Delta}_{\alpha\sigma\sigma^\prime}^* (\delta)c_{i\alpha\sigma^\prime} c_{i+\delta \alpha\sigma}\nonumber \\
    &&~~~~~~+c_{i+\delta \alpha \sigma }^\dagger c_{i\alpha \sigma^\prime}^\dagger \tilde{\Delta}_{\alpha\sigma\sigma^\prime} (\delta)
    - \tilde{\Delta}_{\alpha\sigma\sigma^\prime}^* (\delta) \tilde{\Delta}_{\alpha\sigma\sigma^\prime}(\delta)
    \bigg), 
\end{eqnarray}
where $\tilde{\Delta}_{\alpha\sigma\sigma^\prime} (\delta)= \langle c_{i\alpha\sigma^\prime} c_{i+\delta\alpha\sigma} \rangle$ is the spatially uniform real-space pairing order parameter on NNN bond $\delta$, as shown in Fig.~\ref{fig:afm}(a). To further determine the ansatz of the pairing order parameter $\tilde{\Delta}_{\alpha\sigma\sigma^\prime} (\delta)$, we analyze possible pairing symmetries, which should be classified by the irreducible representations of the symmetry group. Since the addition of displacement field $\mathcal{V}_z$ preserves time-reversal symmetry $\mathcal{T}$ but breaks pseudo-inversion symmetry $\mathcal{I}$, we only consider the $S_z=0$ sector for the U(1) spin-valley symmetry (i.e. inter-valley pairing). And these $S_z=0$ pairings can be further categorized into singlet pairing $\Delta_{\alpha}^S(\delta)=(\tilde{\Delta}_{\alpha\uparrow\downarrow}(\delta)-\tilde{\Delta}_{\alpha\downarrow\uparrow}(\delta))/\sqrt{2}$ and triplet pairing $\Delta_{\alpha}^T(\delta)=(\tilde{\Delta}_{\alpha\uparrow\downarrow}(\delta)+\tilde{\Delta}_{\alpha\downarrow\uparrow}(\delta))/\sqrt{2}$. For the point group symmetry $C_{3v}$ under finite displacement field,  
it is straightforward to show that only $A_1$ (mixing $s$ and $f$-wave) and $E$ (mixing $(p_x,p_y)$ and $(d_{xy},d_{x^2-y^2})$-wave) representations are possible for NNN pairing. The SC pairing form factors, focusing only on chiral SC in the $E$ representation, are illustrated in Fig.~\ref{fig:afm}(b), and we leave the discussion of nematic SC in the $E$ representation in Sec.~VI of the Supplementary Information~\cite{supp}, where we have shown it is subdominant.

To understand the competing insulating phase, we decouple the onsite Hubbard repulsion $H_U$ in the channel of magnetic ordering in the $xy$-plane~\cite{PhysRevResearch.2.033087, PhysRevB.104.075150} as:
\begin{eqnarray}\label{mfhubbard}
    &&H_U = -\sum_{i\alpha} U_\alpha \left(
       m_{i\alpha} c_{i\alpha\downarrow}^\dagger c_{i\alpha\uparrow} +  m_{i\alpha}^* c_{i\alpha\uparrow}^\dagger c_{i\alpha\downarrow} \right) \nonumber\\
     &&~~~~~~~~~+\sum_{i\alpha}U_\alpha |m_{i\alpha}|^2 + \frac{1}{2} \sum_{i\alpha\sigma}U_\alpha n_{i\alpha\sigma}
\end{eqnarray}
where $m_{i\alpha} = \langle c_{i\alpha\uparrow}^\dagger c_{i\alpha\downarrow} \rangle = \langle S_{i\alpha}^x\rangle + i\langle S_{i\alpha}^y\rangle$ is the complex order parameter representing the in-plane magnetization. We further constraint the form of $m_{i\alpha}$ by noting that, as shown in Fig.~\ref{fig:fermisurface}(a), although the Fermi surface deforms as the displacement field $\mathcal{V}_z$ changes, an approximate nesting between spin up and spin down Fermi surfaces persist. Moreover, the nesting wave vectors are relatively close to the commensurate ones $\pm\bm{Q}=(0, \pm4\pi/3a_M)$ over a wide range of displacement field $\mathcal{V}_z$, giving raise to the $120^\circ$ AFM order $m_{i\alpha} = m_{\alpha}^+ e^{iQ\cdot R_i} + m_{\alpha}^- e^{-iQ\cdot R_i}$ with two possible chiralities as shown in Fig.~\ref{fig:afm}(c), which are generally non-degenerate under finite $\mathcal{V}_z$. Such AFM order breaks translation symmetry of the system, folding the Brillouin zone as shown in Fig.~\ref{fig:afm}(d), with the new filling factor in terms of the folded Brillouin zone becoming $\tilde{\nu} = -3$. Depending on details of the system, the ground state can be either metallic or insulating due to the nesting of $\pm\bm{Q}$ is non-perfect.

\begin{figure*}[h]
\centering
\includegraphics[width=0.9\textwidth]{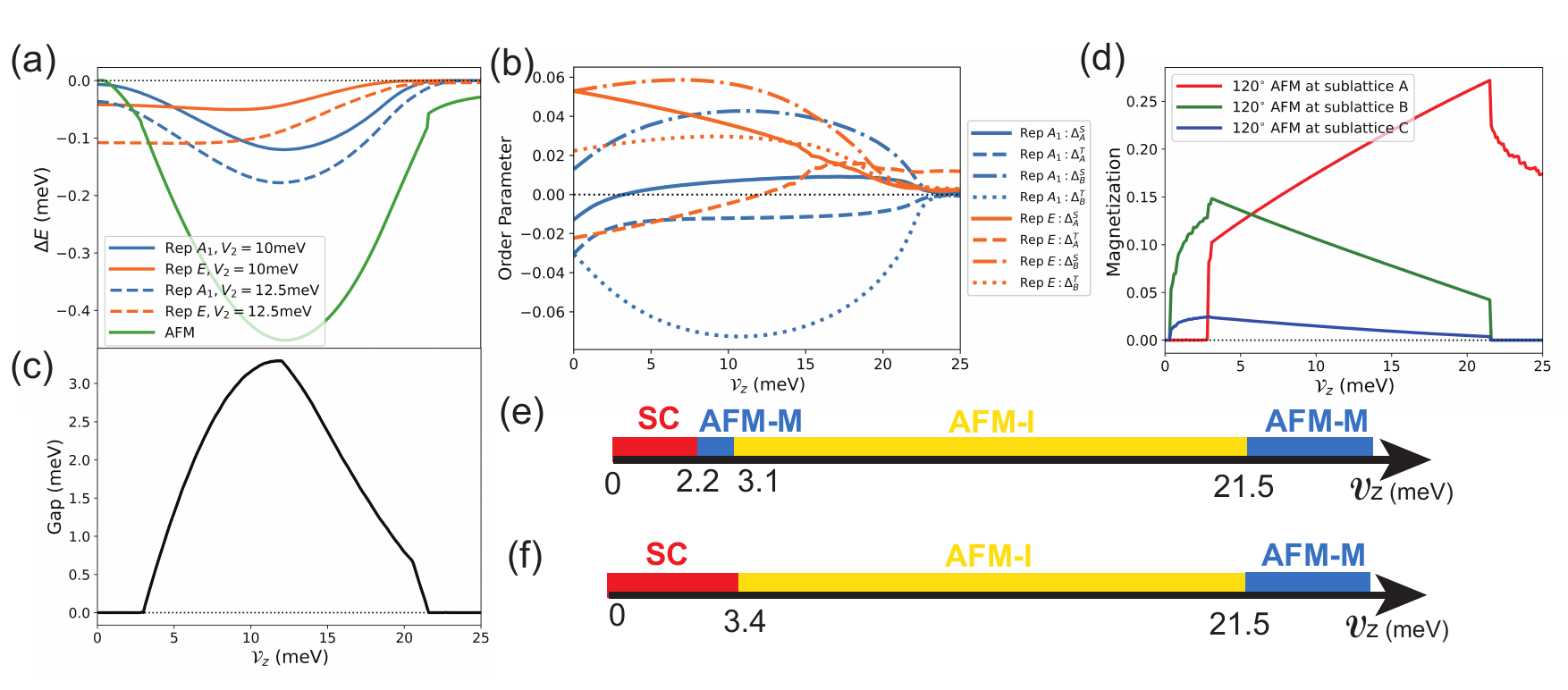}
\caption{\textbf{Mean-field analysis and phase diagrams of $3.65^\circ$ tWSe$_2$.} (a) The energy gain per hole of the ordered phases compared to the symmetric phase. (b) SC orders for $V_2=10$ meV on one specific NNN bond $\delta_0$ under the gauge described in the main text. The sign of order parameters correspond to the sign of real (imaginary) part of singlet (triplet) pairings. (c) The AFM mean-field charge gap at filling factor $\tilde{\nu}=-3$ in the folded Brillouin zone.  (d) Magnitude of the AFM orders on A, B, C sublattices. (e, f) The phase diagrams for $V_2=10$ meV and $V_2=12.5$ meV respectively, indicating that under mean-field framework, the SC to AFM-I transition can either have an intermediate AFM-M phase or occur directly.} \label{fig:phasediagram}
\end{figure*}

We then perform mean-field calculations for SC and AFM at filling $\nu=-1$ independently, using Hubbard repulsion $U_A=U_B=U_C = 37.5$ meV and two different NNN attractions $V_2=10$ meV or $V_2=12.5$ meV as representing parameters. These interacting parameters can be estimated either by comparing the mean-field results with experimental observations, or by directly expanding the gate-screened Coulomb interaction onto Wannier functions~(see Sec.~III and Sec.~IV of the Supplementary Information~\cite{supp}). 
The resulting phase diagrams for tuning the displacement field $\mathcal{V}_z$ at filling $\nu=-1$ are summarized in Figs.~\ref{fig:phasediagram}(e) and (f), respectively. We leave more detailed mean-field derivations and discussions in Sec.~VI and Sec.~VII of the Supplementary Information~\cite{supp}.

\textit{Topological superconductivity.}---We first focus on the SC phase in the zero- or small-displacement field regime. A comparison of the energy gain $\Delta E$ between possible SC orders and the $120^\circ$ AFM order is shown in Fig.~\ref{fig:phasediagram}(a), which indicates that the system is in the SC phase for $\mathcal{V}_z<\mathcal{V}_{z,c}\approx 2.2$ meV ($3.4$ meV)  
for $V_2=10$ meV ($12.5$ meV), qualitatively  
consistent with the critical field $\mathcal{V}_{z,c}^\text{exp}\approx 2.6$ meV observed in the experiment~\cite{xia2024unconventional}. The pairing of this SC phase is  
chiral with mixed $d_{xy}\pm id_{x^2-y^2}$ and $p_x\mp ip_y$ wave in the $E$ representation of $C_{3v}$, spontaneously breaking the time-reversal symmetry $\mathcal{T}$~\cite{cheng2010stable}. More importantly, such a chiral SC phase, which fits into the A-class (namely, breaking $\mathcal{T}$ but preserving the 
$S_z$) of the Altland-Zirnbauer tenfold classification scheme~\cite{PhysRevB.55.1142,schnyder2008classification,chiu2016classification}, is topologically non-trivial with Chern number 
computed to be $C=\pm 2$, suggesting tWSe${}_2$ as a promising candidate for realizing chiral topological SC. The internal structure of the SC order as a function of $\mathcal{V}_z$ for $V_2=10$ meV is also illustrated in Fig.~\ref{fig:phasediagram}(b), where the consistency constraint between time-reversal symmetry $\mathcal{T}$ and $C_{3v}$ mirror plane symmetry enables us to fix a simple gauge such that all singlet pairings $\Delta^S_\alpha(\delta)$ are real while all triplet pairings $\Delta^T_\alpha(\delta)$ are imaginary. It is worth emphasizing that, unlike the single-band moiré Hubbard model~\cite{wang2020correlated,PhysRevResearch.2.033087,PhysRevB.104.075150,10.1063/5.0077901,PhysRevResearch.4.043048,PhysRevResearch.5.L012034,PhysRevLett.130.126001,PhysRevB.106.235135,PhysRevB.108.L201110,PhysRevB.108.155111,PhysRevB.108.064506,akbar2024topological,pan2020quantum, PhysRevB.88.041103,PhysRevB.100.060506}, the mixing of singlet and triplet pairings is allowed for all $\mathcal{V}_z$ in our three-band model, especially for $\mathcal{V}_z=0$, where the emergent spin-valley SU(2) symmetry is absent and the pseudo-inversion symmetry $\mathcal{I}$ does not forbid such mixing. Also, the dominance of the $d_{xy}\pm id_{x^2-y^2}$ component over the $p_x\mp ip_y$ one in the SC phase is in accordance with the result of Chern number $\pm 2$. Moreover, SC is enhanced (suppressed) on the B (A) sublattice upon 
increasing the displacement field $\mathcal{V}_z$ from zero, closely following the trend of the Fermi surface DOS in Fig.~\ref{fig:fermisurface}(b). 

\textit{Antiferromagnetic correlated insulators.}---As the displacement field is further increased beyond a critical field $\mathcal{V}_{z,c}$, 
the system transitions into the $120^\circ$ AFM phase. 
To further determine its transport properties, we 
compute the AFM mean-field charge gap for $\tilde{\nu}=-3$. As shown in Fig.~\ref{fig:phasediagram}(c), 
a broad intermediate range of the AFM insulator (AFM-I) phase appears for 
$\mathcal{V}_{z,c} \lesssim  \mathcal{V}_z < \mathcal{V}^\prime_{z,c}\approx 21.5$ meV and an AFM metal (AFM-M) phase for 
$\mathcal{V}_z>\mathcal{V}^\prime_{z,c}$, closely matching the phenomenology of the experiment~\cite{xia2024unconventional}. 
In the mean-field framework, depending on the value of $V_2$ employed in the model, 
the SC to AFM-I transition either exhibits a tiny intermediate AFM-M phase as shown in Fig.~\ref{fig:phasediagram}(e) or occurs as a direct first-order transition as shown in Fig.~\ref{fig:phasediagram}(f). To explain the transport 
evidence of continuous superconductor-insulator transition~\cite{xia2024unconventional}, disorder could play an important role. The continuous transition into the SC phase might potentially be 
a percolation transition~\cite{Deutscher1984,PhysRevB.27.1541} of local SC regions induced by disorders or a disorder-rounded first-order transition. 
We also present the magnitude of AFM orders on different sublattices as a function of $\mathcal{V}_z$ in Fig.~\ref{fig:phasediagram}(d). Except for small $\mathcal{V}_z$, where the AFM order is stronger on the B sublattice due to its larger DOS (see Fig.~\ref{fig:fermisurface}(b)), the AFM order generally favors the A sublattice, as the holes are mostly concentrated there in response to the displacement field $\mathcal{V}_z$, as well as its better approximation for the commensurate nesting wave vector $\bm{Q}$. And the sudden drop of AFM orders at large $\mathcal{V}_z$ regime coincides with the disappearance of B sublattice hole pockets at $\kappa_\pm$ points and the sharp decline in B sublattice DOS, as illustrated in Fig.~\ref{fig:fermisurface}, indicating that the holes on B sublattice play a crucial role in mediating the AFM order.

\section{Discussion}
In summary, we have constructed a three-band tight-binding model through direct Wannierization, and incorporated onsite Hubbard repulsion and NNN attraction to explain the SC and correlated insulator phase observed in the $3.65^\circ$ tWSe${}_2$ at filling $\nu=-1$~\cite{xia2024unconventional}. Our mean-field analysis indicates that, the SC phase is an A-class topological SC with Chern number $C=\pm 2$, featuring the inter-valley mixed $d_{xy}\pm id_{x^2-y^2}$ and $p_x\mp ip_y$-wave pairing symmetry, and the correlated insulator phase is explained by the $120^\circ$ AFM order. We further demonstrate that the metallic behavior when the filling is away from $\nu=-1$ can also be qualitatively understood within our mean-field framework (see Sec.~VIII of the Supplementary Information~\cite{supp}). 
Compared with the $5^\circ$ tWSe$_2$ system~\cite{guo2024superconductivitytwistedbilayerwse2}, the experimental features show notable similarities despite differences in low-energy band structure and interaction strength~\cite{crepel2024bridging}. Recent theoretical advances~\cite{fischer2024theory} suggest the possibility of a unifying underlying mechanism, motivating future efforts toward a unified description.

The topological band structure in our model can be crucial for understanding how SC arises for flat band systems, since non-trivial lower bounds of SC superfluid weight exist~\cite{peotta2015superfluidity,julku2016geometric,liang2017band,PhysRevLett.128.087002,huhtinen2022revisiting,torma2022superconductivity} due to quantum geometric effects, which deserves more detailed future theoretical studies. Moreover, our results also suggest that the tWSe${}_2$ homobilayers could provide new possibilities for realizing topological SC, which call for 
more detailed experimental studies to further uncover its nature. Identifying the topological edge states by measuring quantized thermal Hall conductance~\cite{kasahara2018majorana,banerjee2017observed,banerjee2018observation}, or verifying the chiral nature of the SC pairing symmetry through the phase-sensitive Josephson junction~\cite{PhysRevLett.71.2134,van1995phase,tsuei2000pairing}, will surely open new opportunities in twisted TMD systems. 

\textit{Note added.}---In finishing the present work, we noticed that Refs. \cite{kim2024theory, myersonjain2024superconductorinsulatortransitiontmdmoire,zhu2024theory, christos2024approximatesymmetriesinsulatorssuperconductivity,xie2024superconductivitytwistedwse2topologyinduced,guerci2024topologicalsuperconductivityrepulsiveinteractions} appeared, which also investigated the SC and correlated insulators in tWSe${}_2$ 
reported in Ref.~\cite{xia2024unconventional}, although there are important differences between those studies and the present one. 

\section{Methods}
We construct a three-band tight-binding model by a direct Wannierization of the continuum model of $3.65^\circ$ twisted bilayer WSe2, with the resulting hopping parameters given in Supplementary Information Sec.~I. The model incorporates onsite Hubbard repulsion and nearest- or next-nearest-neighbor attractions. A detailed estimation of the Hubbard interaction parameters is provided in Supplementary Information Sec.~III and Sec.~IV, while a microscopic explanation of the attractive terms is presented by the random phase approximation calculation in Supplementary Information Sec.~V. The interacting model is analyzed within the standard mean-field framework, with full derivations and self-consistent procedures provided in Supplementary Information Sec.~VI and Sec.~VII.

\section{Data Availability}
Source data for all figures in the main article are available in the Supplementary Data file.

\section{Code Availability}
The codes used in this study are available from the corresponding author upon request.

\section{References}
\bibliography{ref.bib}

\section{Acknowledgements}
We would like to thank Andrei Bernevig 
for helpful discussions. 
This work is supported in part 
by MOSTC under Grant No. 2021YFA1400100 (H.Y.), 
by the Innovation Program for Quantum Science and Technology under Grant No.2021ZD0302502 (H.Y.),  
by NSFC under Grant Nos. 12347107 (C.T., M.-R.L., Z.W., W.S., H.Y.) and 12334003 (H.Y.), 
and by the New Cornerstone Science Foundation through the Xplorer Prize (H.Y.).

\section{Author Contributions Statement}
H.Y. conceived and supervised the project. C.T. and M.-R.L. carried out the computations. All authors (C.T., M.-R.L., Z.W., W.S., and H.Y.) contributed to the interpretation of results and to writing the manuscript.

\section{Competing Interests Statement}
The authors declare no competing interests.

\section{Figure Legends/Captions}
\begin{figure}[h] 
\centering 
\includegraphics[width=0.5\textwidth]{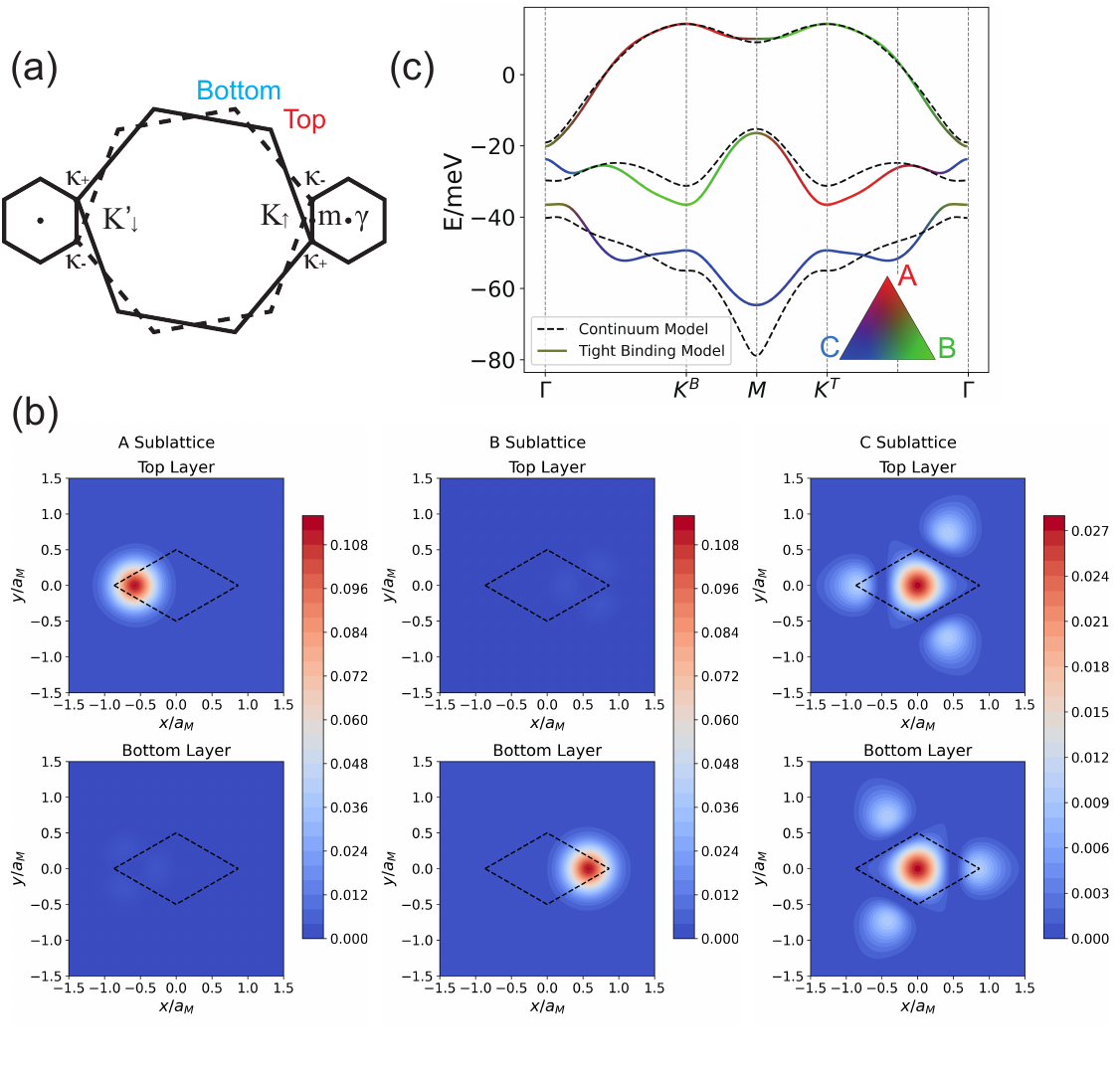} 
\caption{\textbf{Construction of the three-band tight-binding model from the continuum model by Wannierization.} (a) Mini Brillouin zone formed by a small twist angle $\theta$ between two layers. (b) Density distribution of Wannier functions on two layers, with unit cell shown as dashed line. (c) Comparison of the band structure between continuum model (dashed line) and tight binding model (solid line), where colors representing sublattice components. }
\label{fig:fitband}
\end{figure}

\begin{figure}[h]
\centering 
\includegraphics[width=0.48\textwidth]{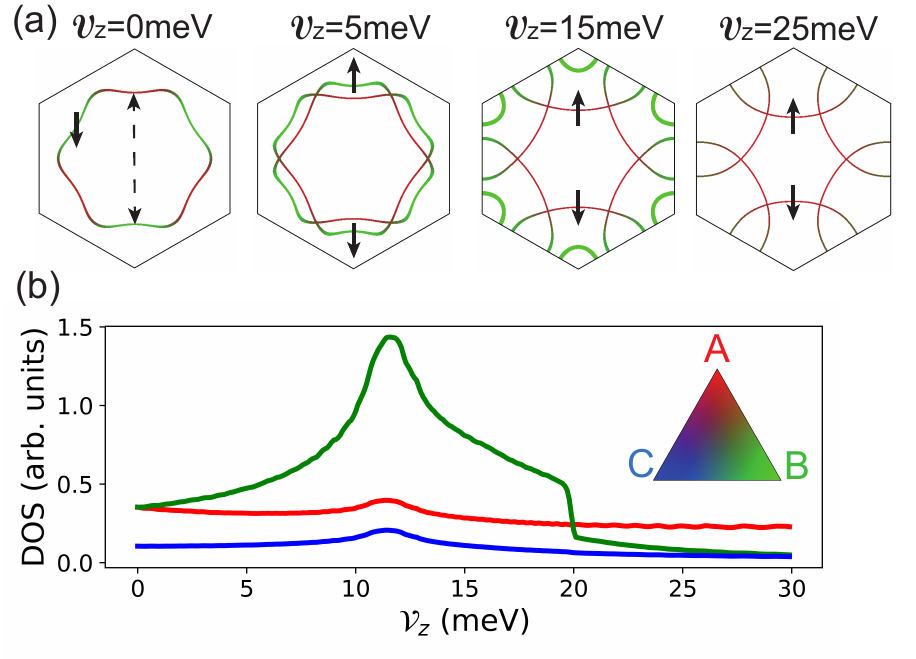} 
\caption{\textbf{Evolution of Fermi surfaces and density of states as a function of displacement field.} (a) Fermi surfaces of free Hamiltonian at different displacement fields, with colors indicating the sublattice components, and thickness representing the DOS. The spin of the Fermi surfaces are labeled as solid arrows. When $\mathcal{V}_z=0$ meV, where spin up and down Fermi surfaces coincide, only spin down Fermi surface is shown. The approximate nesting wave vector $\bm{Q}$ between spin up and down Fermi surfaces are illustrated as dashed arrow. (b) The Fermi surface DOS of A, B, C sublattices as a function of displacement field $\mathcal{V}_z$.}
\label{fig:fermisurface}
\end{figure}

\begin{figure}[h]
\centering
\includegraphics[width=8cm]{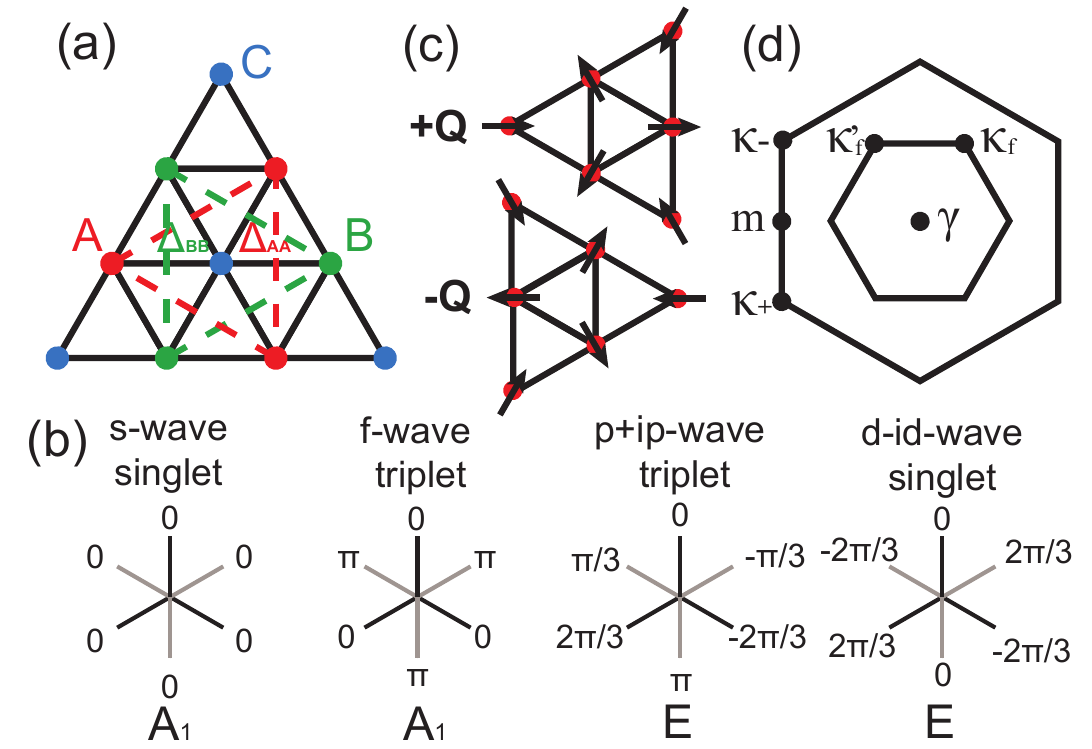}
\caption{\textbf{Illustration of the possible superconducting and antiferromagnetic order parameters.} (a) The NNN SC order parameters $\Delta_{AA}$ and $\Delta_{BB}$, where the A, B, C sites are labeled as red, green, blue dots. (b) The pairing form factor of $s$-wave, $f$-wave, $p+ip$-wave, and $d-id$-wave on NNN bonds, with the irreducible representation labeled. (c) $120^{\circ}$ AFM pattern on a certain type of sublattice with wave vector $\pm Q$. (d) The folded Brillouin zone induced by the AFM order. } \label{fig:afm}
\end{figure}

\onecolumngrid

\clearpage
\widetext

\begin{center}
\textbf{\large Supplemental Material for ‘Theory of Topological Superconductivity and Antiferromagnetic Correlated Insulators in Twisted Bilayer WSe${}_2$’}
\end{center}

\renewcommand{\thefigure}{S\arabic{figure}}
\renewcommand{\theHfigure}{S\arabic{figure}}
\setcounter{figure}{0}
\renewcommand{\theequation}{S\arabic{equation}}
\setcounter{equation}{0}
\renewcommand{\thetable}{S\arabic{table}}
\setcounter{table}{0}
\renewcommand{\thesection}{\Roman{section}}
\setcounter{section}{0}
\setcounter{secnumdepth}{4}

\section{Hopping Parameters of the Tight-Binding Model}\label{supp:hopping}
\begin{figure}[h] 
\centering 
\includegraphics[width=0.5\textwidth]{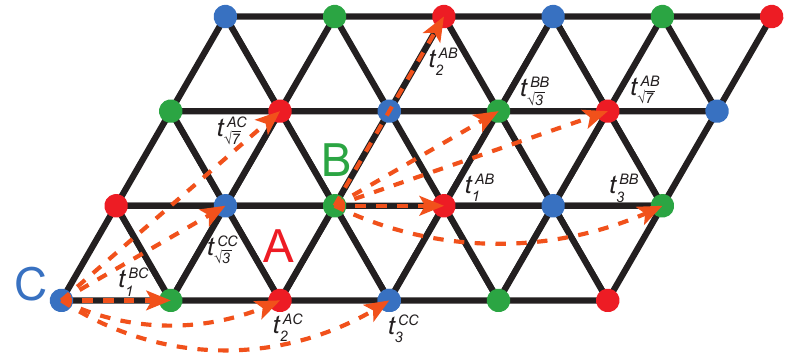} 
\caption{Illustration of the representative hopping bonds up to $5$th-nearest-neighbor.}
\label{fig:hopping}
\end{figure}

In this section, we provide additional information of the tight-binding model hopping parameters we adopted in the main text, derived from direct Wannierization of the continuum model for $3.65^\circ$ tWSe${}_2$ at $V_z=0$. Fig.~\ref{fig:hopping} illustrates the symmetry inequivalent representative hopping bonds up to $5$th-nearest-neighbor, and we provide the hopping parameters for $K$ valley  on these representative bonds as well as the onsite potentials as:
\begin{equation}
\begin{aligned}
    &t_0^A = t_0^B = -15.26 \text{ meV}, \quad t_0^C = -42.89 \text{ meV} \\
    &t^{BC}_1 = 7.12\text{ meV}, \quad t^{CC}_{\sqrt{3}} = 2.76 \text{ meV}, \quad t^{AC}_2 = 1.15 \text{ meV} \\
    &t^{AC}_{\sqrt{7}} = 3.01 e^{i 0.965\pi}\text{ meV}, \quad t^{CC}_3 = 0.43 \text{ meV} \\
    &t^{AB}_1 = -4.99 \text{ meV}, \quad t^{BB}_{\sqrt{3}} = 5.16 e^{i 0.618 \pi} \text{ meV}\\
    & t^{AB}_2 = 0.7 \text{ meV}, \quad t^{AB}_{\sqrt{7}} = 0.78 \text{ meV}, \quad t^{BB}_3 = -0.31 \text{ meV}
\end{aligned}
\end{equation}
where $t_d^{\alpha\beta}$ represents the hopping parameter from $\beta$ sublattice to $\alpha$ sublattice with distance $d$ (set NN with $d=1$), and $t_0^\alpha$ denotes the onsite potential of $\alpha$ sublattice.

There are several remarks on the symmetry properties of these hopping parameters. First, hopping parameters related by $C_{3z}$ symmetry will (will not) be identical if $\alpha, \beta$ have identical (different) $C_{3z}$ eigenvalues (see main text). Second, $C_{2y}\mathcal{T}$ and $\mathcal{I}$ symmetries constrain most but not all hopping parameters to be real. Third, the hopping parameters for $K$ and $-K$ valley are related by time-reversal symmetry $\mathcal{T}$. Last, unlike the single-band moiré Hubbard model, where emergent spin-valley SU(2) symmetry exists at $V_z=0$, the spin-valley symmetry here remains U(1) due to the presence of complex hopping parameters.

\section{Displacement Field Dependence of the Wannier Functions}\label{supp:wannier_d}

\begin{figure}[h] 
\centering 
\includegraphics[width=0.8\textwidth]{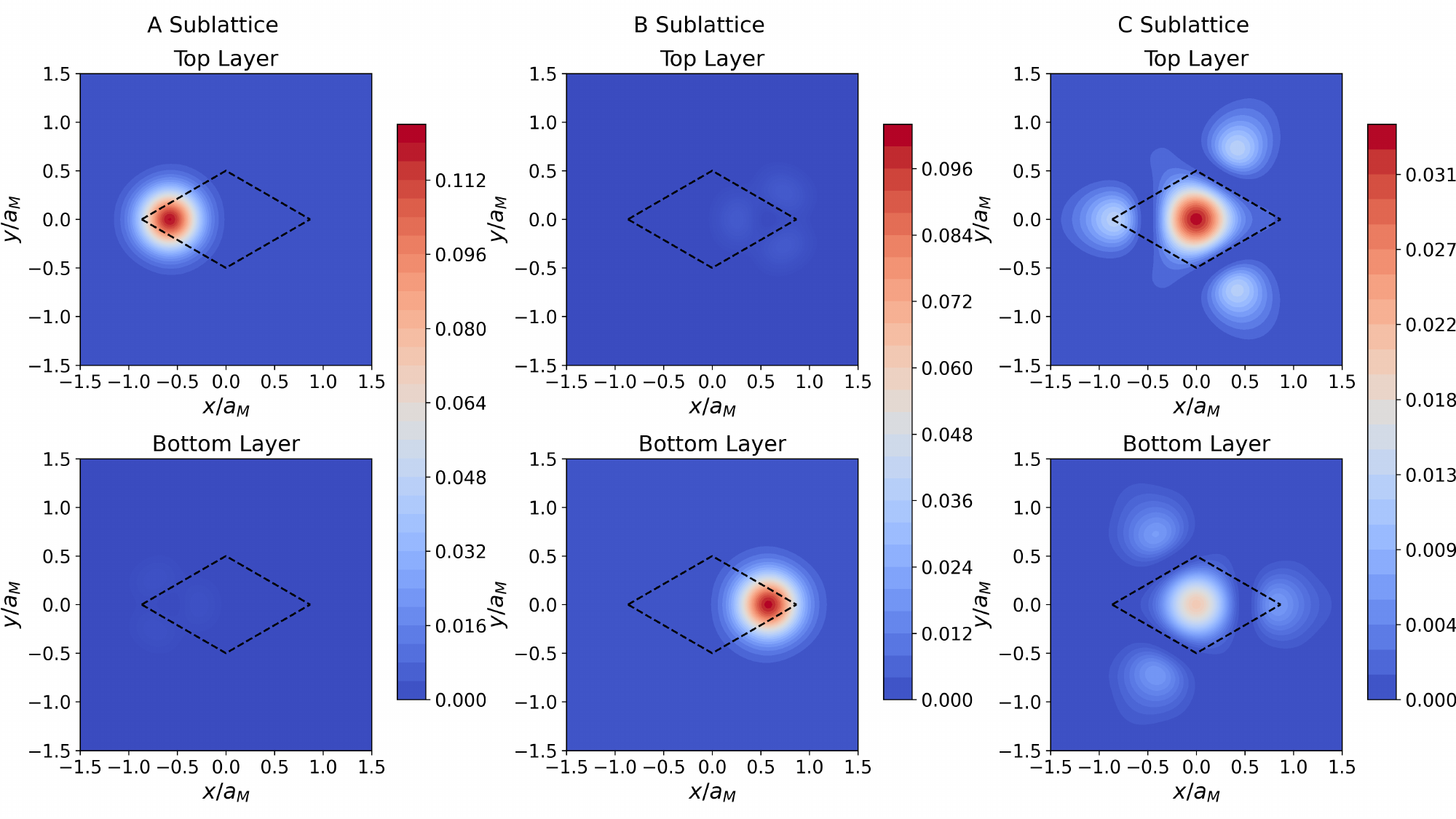} 
\caption{Density distribution of Wannier functions on two layers at $V_z=20$ meV, with unit cell shown as dashed line.}
\label{fig:wannier_d}
\end{figure}

The layer-polarization Wannierization scheme can be applied not only to tWSe$_2$ at zero displacement field, but also at finite displacement fields. Fig.~\ref{fig:wannier_d} illustrates the density distribution of Wannier functions obtained from the continuum model with the displacement field set to $V_z = 20$ meV (see Eq.~\eqref{moire_potential_and_tunneling} in the main text).

Compared with the Wannier functions at zero displacement field (Fig.~\ref{fig:fitband}(b) in the main text), we find that the A and B Wannier functions remain almost unchanged (A (B) becomes slightly more localized (extended)) and remain strongly polarized in opposite layers. In contrast, the C Wannier function exhibits a more noticeable redistribution though remaining layer hybridized. This behavior can be simply understood as follows: the A and B Wannier functions are approximate eigenstates of the displacement field due to their strong layer polarization, while the C Wannier function is not. Nevertheless, we neglect this modification of the C orbital in the main text, as it lies far from the Fermi surface and does not play a significant role in the low-energy physics.

To further justify this approximation, we compute the onsite chemical potential terms $t_0^\alpha$ of the Wannier functions at $V_z = 20$ meV: $t_0^A = -7.77$ meV, $t_0^B = -22.72$ meV, and $t_0^C = -41.10$ meV. These results confirm that the C orbital remains energetically well separated from the low-energy physics. Moreover, we can extract the effective displacement field $\mathcal{V}_z = t_0^A-t_0^B = 14.95$ meV in our tight-binding model (see Eq.~\eqref{hd} in the main text). The fact that $\mathcal{V}_z$ is slightly smaller than the original $V_z$ can be attributed to the A and B Wannier functions not being perfectly layer polarized. These analyses confirm the robustness of the assumptions underlying our model.

\section{Mean-field Results with Different Interacting Parameters}\label{supp:mf_diff_para}
In the main text, we adopt the interacting parameters $U_A=U_B=U_C=37.5$ meV, $V_2 = 10$ meV or $12.5$ meV as representative parameters of $3.65^\circ$ tWSe$_2$. This section further explores our mean-field model over a broad range of interacting parameters, offering valuable physical insights into the tWSe$_2$ system and providing justifications of the parameter choices.

\begin{figure}[h] 
\centering 
\includegraphics[width=0.8\textwidth]{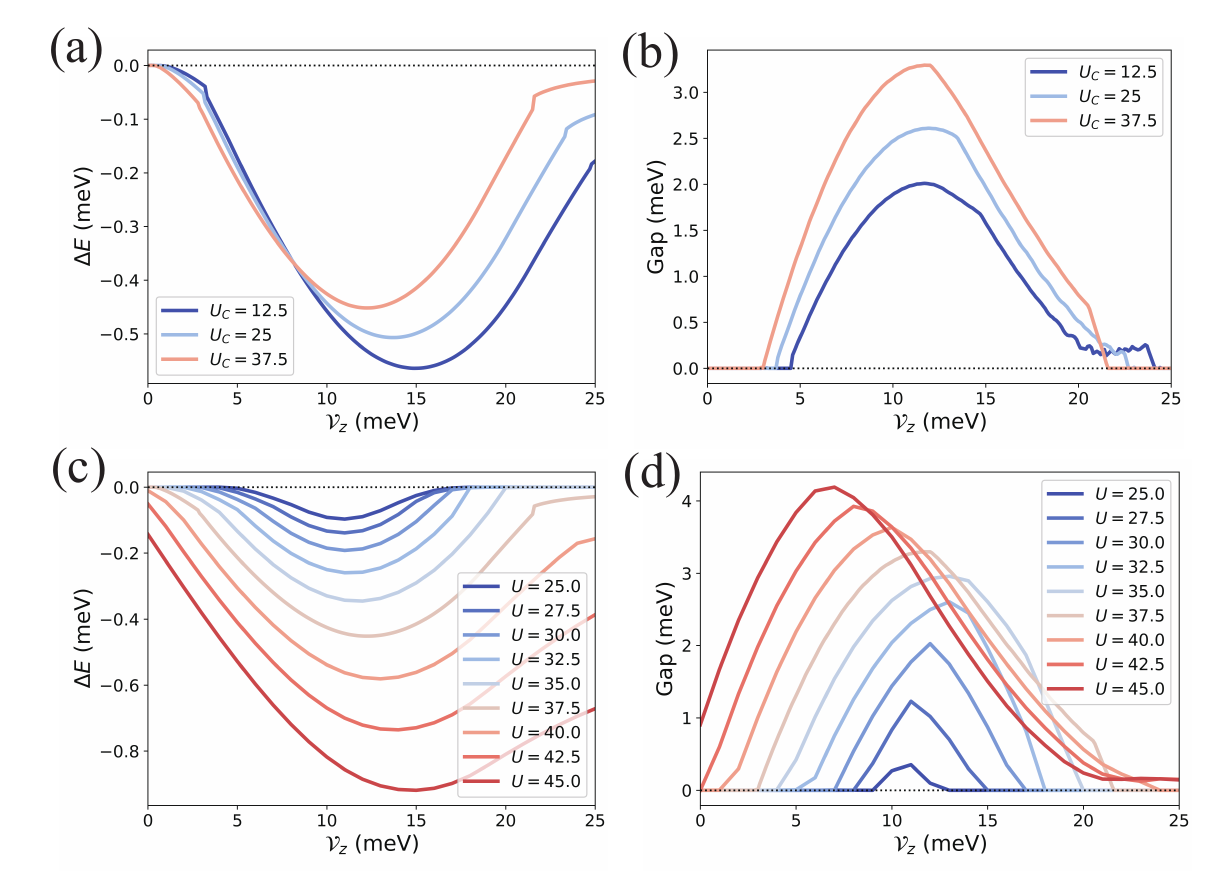} 
\caption{(a,b) The AFM energy gain per hole and the mean-field charge gap with $U_{A}=U_{B}=37.5$ meV and different $U_C$. (c,d) The AFM energy gain per hole and the mean-field charge gap with various $U_{A}=U_{B}=U_{C}=U$.}
\label{fig:changing_U}
\end{figure}

We start with the discussion of onsite Hubbard interaction. We emphasize that, although symmetries only impose $U_A = U_B$ while allowing $U_C$ to differ, the parameter choice can be simplified by setting $U_A = U_B = U_C = U$, as the C orbital is far from the Fermi surface (see Fig.~\ref{fig:fermisurface} in the main text). Such simplification reduces independent tuning parameters without altering the essential physical properties, making the qualitative physics more transparent. Fig.~\ref{fig:changing_U} (a,b) illustrates the AFM energy gain per hole and the mean-field charge gap with $U_A=U_B=37.5$ meV with varying $U_C$, showing only minor differences (especially for the small displacement field regime close to the SC phase) even when $U_C$ is tuned over a relatively large range. 
To gain further insights into the strength of the Hubbard interaction, we then keep $U_A=U_B=U_C=U$ and vary the Hubbard $U$ from $25$ meV to $45$ meV. The results are shown in Fig.~\ref{fig:changing_U} (c,d), where both the AFM energy gain and the charge gap increase with $U$ as expected. Based on these results, we conclude that $U$ in the range of $35 \sim 40$ meV is appropriate, where smaller $U$ weakens the insulating behavior while larger $U$ stabilizes AFM ground state over SC for small displacement field.

We now briefly comment on the values of the NNN attraction $V_2$. While an infinitesimal attraction $V_2$ should be sufficient to induce SC tendency (due to the Cooper instability), a relatively large $V_2 = 10$ or $12.5$ meV is adopted in the main text. This choice aims to provide a better explanation for the continuous SC-insulator transition in the experiment. In the main text, we attribute such continuous transition to disorder, which requires our mean-field results (without disorder) either exhibit a weak first-order direct SC-AFM insulator transition ($V_2=12.5$ meV), or feature a small intermediate AFM metal phase ($V_2=10$ meV). A weaker $V_2$ results in extended intermediate phase, while a larger $V_2$ will cause strong direct first-order transition, which is challenging to make continuous even when disorder is considered.

\section{Interacting Parameters Directly from Wannier Functions}\label{supp:interaction_from_wannier}

In addition to determining the interacting parameters by comparing the mean-field results with the experiment, these parameters can also be directly obtained from the Wannier functions. Given the considerable uncertainties in the microscopic mechanism of attractions, here we primarily focus on the more well-understood dual-gate screened Coulomb repulsion:
\begin{equation}\label{gate_screened_coulomb}
\begin{split}
    H_\text{dual-gate} = \frac{1}{2A} \sum_{k,k^\prime,q,l,l^\prime,\tau,\tau^\prime} V_0(q)& \, c^\dagger_{k+q l\tau} c^\dagger_{k^\prime-q l^\prime\tau^\prime} c_{k^\prime l^\prime\tau^\prime} c_{kl\tau} \\
    V_0(q) = \pi \xi^2 V_\xi \frac{\tanh(\xi |q|/2)}{\xi|q|/2},&\quad V_\xi = \frac{e^2}{4\pi\epsilon_0\epsilon_r \xi}
\end{split}
\end{equation}
where $c_{kl\tau}$ annihilates a plane wave state with momentum $k$ in layer $l$ and spin/valley $\tau$, $A$ denotes the total area of the system, $\xi$ is the distance between the two metal gates, $e$ is the elementary charge, and $\epsilon_0$ ($\epsilon_r$) correspond to the vacuum (relative) dielectric constant.

\begin{figure}[t] 
\centering 
\includegraphics[width=0.6\textwidth]{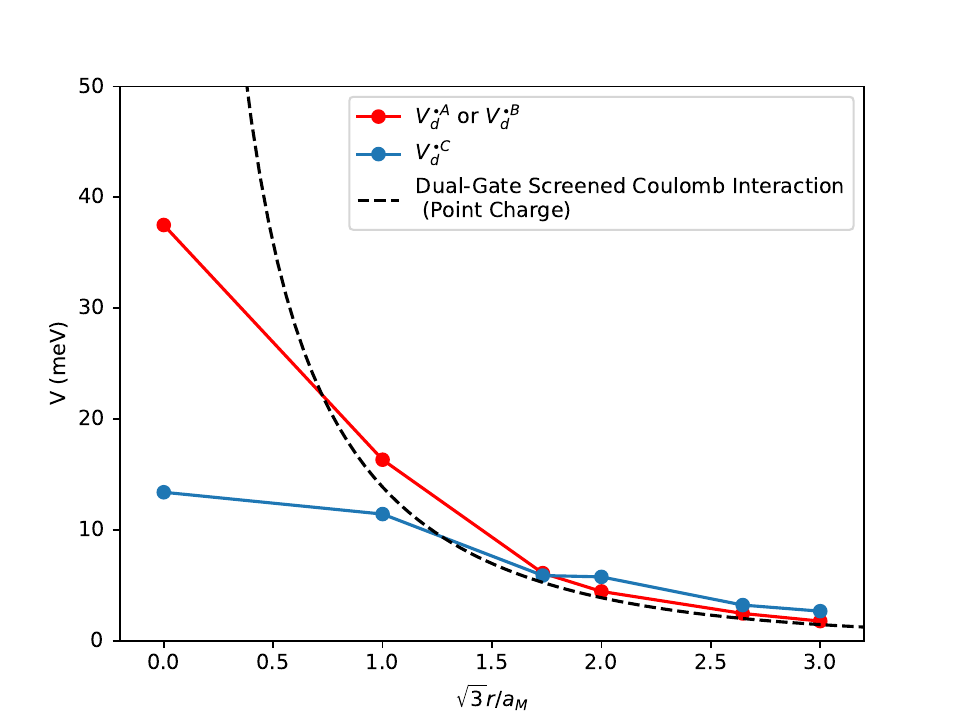} 
\caption{Interacting parameters obtained by expanding the dual-gate screened Coulomb interaction with $\xi=10$ nm onto the Wannier functions. The black dashed line represents the dual-gate screened Coulomb interaction for two point charges.}
\label{fig:interaction}
\end{figure}

Fig.~\ref{fig:interaction} illustrates the interacting parameters obtained by expanding the dual-gate screened Coulomb interaction onto the Wannier functions, using an experimentally relevant gate distance of $\xi=10$ nm. Here, $V^{\bullet \alpha}_d$ denotes the interacting parameter between a Wannier function on $\alpha$ sublattice and another Wannier function at distance $d$ apart on the same representative bond shown in Fig.~\ref{fig:hopping} (set NN with $d=1$, and $d=0$ corresponds to the onsite Hubbard interaction $U_\alpha$). The dielectric constant $\epsilon_r$ is determined by setting $U_{A/B} = 37.5$ meV. We also plot the dual-gate screened Coulomb interaction in real space $V(r)$ (i.e., the Fourier transformation of $V_0(q)$), where the interacting parameters closely follow its trend at large distances. $V(r)$ can be written in summation form as:
\begin{equation}
    V(r) = V_\xi \sum_{n=-\infty}^{+\infty} \frac{(-1)^n}{\sqrt{(r/\xi)^2+n^2}}
\end{equation}
which can be physically interpreted as the sum of all contributions from image charges induced by two metal gates. 

Here, we offer some additional remarks on these interacting parameters. First, to ensure our mean-field results match the experiment, the dielectric constant is determined to be $\epsilon_r \sim 21.4$ (for $U_{A/B}=37.5$ meV, as in the main text). The relatively large value of $\epsilon_r$ can be attributed to additional screening effects in tWSe$_2$ systems beyond dual-gate screening, including contributions from the dielectric environment or higher-energy bands. Second, we still adopt $U_A=U_B=U_C$ in the main text, though the Hubbard interaction $U_C$ is considerably smaller than $U_{A/B}$ in Fig.~\ref{fig:interaction} (which can be naturally understood by the extension of the Wannier functions, see Fig.~\ref{fig:fitband}(b) in the main text). The main reason is that the Wannier functions obtained by the layer-polarization Wannierization scheme are known to be not maximally localized. In particular, the (not so localized) C Wannier function will become more localized if we further optimize its spread functional, resulting in a significantly larger $U_C$ (closer to $U_{A/B}$). Moreover, as already illustrated in Fig.~\ref{fig:changing_U}(a,b), changing $U_C$ does not affect the main physics of the system, since the C Wannier function is away from the Fermi surface. Third, it is reasonable to retain only the onsite Hubbard interaction while neglecting further neighbor interactions. As shown in Fig.~\ref{fig:interaction}, interactions beyond NN are negligibly small, allowing us to safely discard them. Although the NN interaction (from directly expanding the dual-gate Coulomb interaction) has considerable strength (with $V^{AB}_1 / U_{A/B} \sim 0.44$), we argue that the effective $V_1$ should be significantly reduced when accounting for possible attraction mechanisms or additional screening effects beyond dual-gate screening. Moreover, further minimization of the Wannier function spread can further enhance the Hubbard $U$. Therefore, our model retaining only onsite Hubbard repulsion should provide a good starting point for understanding the physical properties of tWSe$_2$.

\section{
Attractive Interactions from Random Phase Approximation Analysis}\label{supp:rpa}
In this section, we provide a microscopic justification of the phenomenological attraction adopted in the main text, using a random phase approximation (RPA) analysis. The RPA method captures collective screening effects by summing the infinite series of bubble diagrams in the particle-hole channel, which can result in effective attractive interactions in certain real-space regions. In moiré superlattice systems, due to the small size of moiré Brillouin zone, Umklapp scattering must be treated explicitly in the RPA calculation, as we detailed below. 

Here, we consider the RPA screening effects of dual-gate screened Coulomb interaction in Eq.~\eqref{gate_screened_coulomb} projected to the topmost moiré band. The band projected density operator $\rho_{q+g}$ (where $q$ is in moiré Brillouin zone, $g$ is reciprocal wave vector) is given by:
\begin{equation}\label{pdo}
    \rho_{q+g} = \sum_{k,\sigma} \Lambda^{k+q,k}_{g\sigma} c^\dagger_{k+q,\sigma} c_{k,\sigma}
\end{equation}
where $\Lambda^{k+q,k}_{g\sigma} $ is the form factor of the projected density operator, defined by:
\begin{equation}
    \Lambda^{k_1,k_2}_{g\sigma} = \sum_{g^\prime, l} z^*_{g+g^\prime,l,\sigma}(k_1) z_{g^\prime,l,\sigma}(k_2) 
\end{equation}
with the definition of $z_{g,l,\sigma}(k)=\langle g,l|u_{k,\sigma}\rangle$, where $\ket{g,l}$ is the plane wave state with momentum $g$ at layer $l$, and $\ket{u_{k\sigma}}$ is the periodic part of the Bloch state with momentum $k$ and spin/valley $\sigma$ for the topmost moiré band.

There are several symmetry properties of the form factor $\Lambda^{k_1,k_2}_{g\sigma}$ that can be useful:
\begin{equation}\label{lambda_trans}
    \Lambda^{k_1,k_2}_{g\sigma} = \Lambda^{k_1 + g^\prime,k_2}_{g-g^\prime\sigma} = \Lambda^{k_1 ,k_2+g^\prime}_{g+g^\prime\sigma} 
\end{equation}
\begin{equation}\label{lambda_trs}
    \Lambda^{k_1,k_2}_{g\downarrow} = \left[ \Lambda^{-k_1,-k_2}_{-g\uparrow}\right]^* 
\end{equation}
\begin{equation}
    \Lambda^{k_1,k_2}_{g\sigma} = \left[ \Lambda^{k_2,k_1}_{-g\sigma}\right]^*
\end{equation}
where, Eq.~\eqref{lambda_trans} requires the periodic gauge $z_g(k+g^\prime) = z_{g+g^\prime}(k)$, and it is particularly useful for relating arbitrary momentum $k_1, k_2$ back to the first Brillouin zone. Eq.~\eqref{lambda_trs} requires the time-reversal symmetric gauge $z_\downarrow(k)=z_\uparrow^*(-k)$, which can relate the spin down quantities to the corresponding spin up ones.

Having projected the density operator to the topmost band, the particle-hole susceptibility $\Pi(i\omega_m, q)_{g,g^\prime}$ is defined by the time-ordered density correlation function:
\begin{equation}
\begin{split}
    \Pi(i\omega_m, q)_{g,g^\prime} &= \frac{1}{A}\int_0^\beta d\tau e^{i\omega_m \tau} \int dr dr^\prime e^{-i(q+g)r} e^{i(q+g^\prime)r^\prime} \Pi(r,r^\prime,\tau) \\
    &= -\frac{1}{A}\int_0^\beta d\tau e^{i\omega_m \tau} \langle T_\tau\rho_{q+g}(\tau) \rho_{-q-g^\prime}(0)\rangle 
\end{split}    
\end{equation}

Since there are only discrete lattice translation symmetries $\Pi(r+a, r'+a,\tau)=\Pi(r,r',\tau)$ (where $a$ is some moiré lattice vector), momentum is only conserved modulo $g$. The above equation treats the Umklapp scattering structure explicitly and exactly.

Substitute the projected density operator Eq.~\eqref{pdo} into the above equation, we can evaluate the particle-hole susceptibility exactly:
\begin{equation}
\begin{split}
    \Pi(i\omega_m, q)_{g,g^\prime} 
    &= -\frac{1}{A}\int_0^\beta d\tau e^{i\omega_m \tau} \sum_{kk^\prime\sigma\sigma^\prime} \Lambda^{k+q,k}_{g\sigma} \Lambda^{k^\prime-q,k^\prime}_{-g^\prime\sigma^\prime} \langle T_\tau c^\dagger_{k+q,\sigma}(\tau) c_{k,\sigma}(\tau) c^\dagger_{k^\prime-q,\sigma^\prime}(0) c_{k^\prime,\sigma^\prime}(0)\rangle \\
    &= \frac{1}{A}\int_0^\beta d\tau e^{i\omega_m \tau} \sum_{k\sigma} \Lambda^{k+q,k}_{g\sigma} \Lambda^{k,k+q}_{-g^\prime\sigma} G_\sigma(k,\tau)G_\sigma(k+q,-\tau) \\
    &=\frac{1}{A} \sum_{k\sigma} \frac{n_{k\sigma}-n_{k+q\sigma}}{i\omega+\epsilon_{k\sigma}-\epsilon_{k+q\sigma}} \Lambda^{k+q,k}_{g\sigma} \Lambda^{k,k+q}_{-g^\prime\sigma} \\ 
    &=\frac{1}{A} \sum_{k\sigma} \frac{n_{k\sigma}-n_{k+q\sigma}}{i\omega+\epsilon_{k\sigma}-\epsilon_{k+q\sigma}} \Lambda^{k+q,k}_{g\sigma} \left[\Lambda^{k+q,k}_{g^\prime\sigma}\right]^*
\end{split}
\end{equation}
where $G_\sigma(k,\tau) = -\langle T_\tau c_\sigma(\tau) c^\dagger_\sigma(0)\rangle$ is the single particle green's function, $\epsilon_{k\sigma}$ is the band dispersion and $n_{k\sigma}$ is the Fermi-Dirac distribution.

Defining the bare interaction $V_0$ as a diagonal matrix of $g$:
\begin{equation}
    V_0(q)_{g,g^\prime} = V_0(q+g) \delta_{g,g^\prime}
\end{equation}

Then, the RPA screened interaction follows directly from the standard Dyson series summation (summation over bubble diagrams), where all the quantities should be understood as matrix with index $g,g^\prime$:
\begin{equation}
\begin{split}
    V^{RPA}(i\omega_m,q) &= V_0(q) + V_0(q) \Pi(i\omega_m,q) V_0(q) + V_0(q) \Pi(i\omega_m,q) V_0(q) \Pi(i\omega_m,q) V_0(q) + \cdots \\    
    &= [1-V_0(q) \Pi(i\omega_m,q)]^{-1} V_0(q)
\end{split}
\end{equation}

We adopt $V^{RPA}(i\omega_m=0,q)$ as the RPA effective interaction. We can then Fourier transform to the real space:
\begin{equation}
\begin{split}
    V^{RPA}(r,r^\prime) &= \frac{1}{A}\sum_{q,g,g^\prime} V^{RPA}(i\omega_m=0,q)_{g,g^\prime} e^{i(q+g)r} e^{-i(q+g^\prime)r^\prime} \\
    &= \frac{1}{A}\sum_{q,g,g^\prime} V^{RPA}(i\omega_m=0,q)_{g,g^\prime} e^{iq(r-r^\prime)}e^{igr}e^{-ig^\prime r^\prime}
\end{split}
\end{equation}
where it should be emphasize that, the momentum space cutoff of the above Fourier transformation should be taken relatively large, since the dual-gate screened interaction $V_0(q)$ in Eq.~\eqref{gate_screened_coulomb} has slow $1/q$ decay at large $q$.

\begin{figure}[h] 
\centering 
\includegraphics[width=0.95\textwidth]{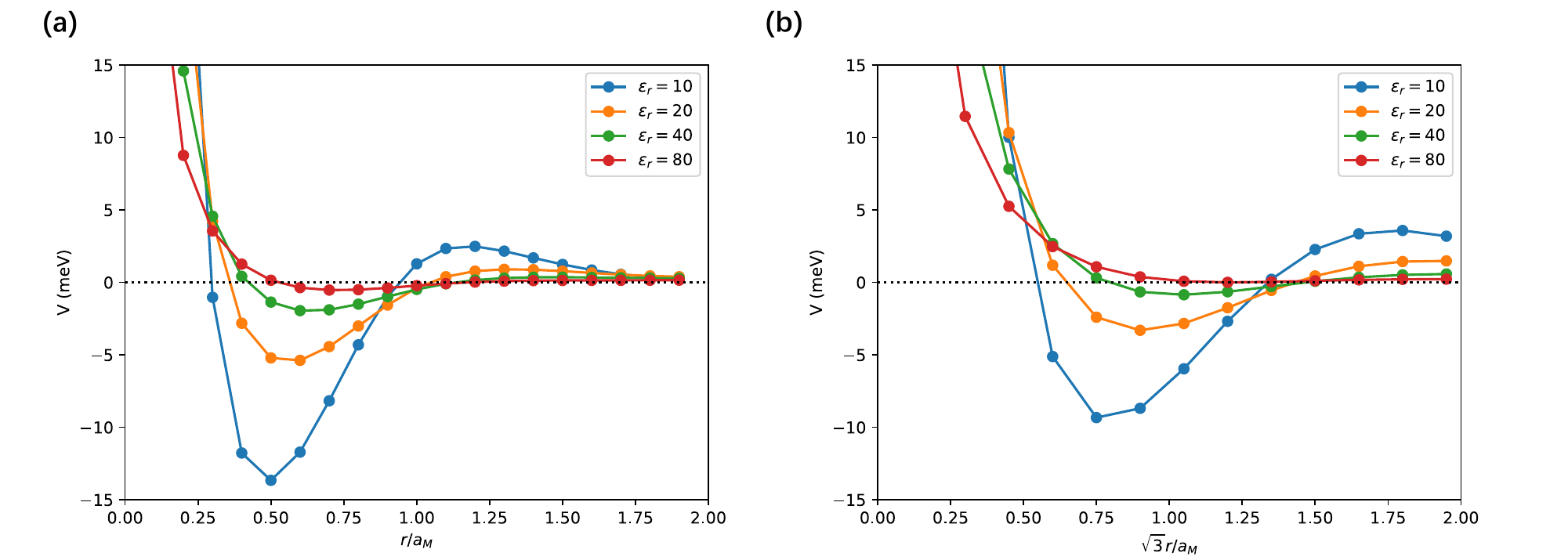} 
\caption{The RPA effective interactions $V^{RPA}(r,r^\prime)$ with $r$ fixed at A sublattice and $r^\prime$ along (a) NNN A-A (b) NN A-B direction with dual-gate distance $\xi=10$ nm and different dielectric constants $\epsilon_r$.}
\label{fig:rpa}
\end{figure}

Fig.~\ref{fig:rpa} illustrates the spatial profile of the RPA effective interaction $V^{RPA}(r,r^\prime)$ along the NNN or NN directions. Here, $r=0$ is located at the Wannier center of A sublattice (i.e. XM region), and the dual-gate distance is fixed at experimental relevant value $\xi=10$ nm. The point $r/a_M=1$ for the NNN direction (Fig.~\ref{fig:rpa}(a)) and the point $\sqrt{3}r/a_M=1$ for the NN direction (Fig.~\ref{fig:rpa}(b)) corresponds to the Wannier center of the NNN or NN site, respectively. These results allow us to evaluate how strongly the Coulomb repulsion is screened and where the effective attraction occurs.

First, we observe that the RPA mechanism further screens the dual-gate screened Coulomb interaction, leading to even more enhanced localization of the repulsive part. This provides further justification of retaining the onsite Hubbard repulsion $H_U$ only in our tight-binding model, as the longer-range repulsions are significantly suppressed.

Moreover, the RPA results reveal that attractive interactions emerge predominantly in the relatively local regions (though some longer-range attractions beyond this range may also occur, but their magnitudes are strongly screened compared with the local ones). Specifically, in Fig.~\ref{fig:rpa}(b), a strong attraction is clearly seen near the NN region $\sqrt{3}r/a_M\sim 1$. While in Fig.~\ref{fig:rpa}(a) the attraction appears in the range $0.4\lesssim r/a_M \lesssim 1$ (at $r/a_M = 1$, only weak attraction occurs for relatively large $\epsilon_r$), the NNN interaction can still be attractive (though weaker than the NN attraction) considering the finite spatial extent of the Wannier function. These findings justify the use of relatively local attractive terms in our tight-binding model. While both NN and NNN attraction are possible, our main text mainly focus on the NNN attraction since, as shown below (Fig.~\ref{fig:energy_gain_supp}), it induces much stronger superconductivity, which can be theoretically understood by the sublattice polarization property of the Fermi surface.

\section{Derivations and Supplemental Results of Superconducting Mean-Field Analysis}\label{supp:mf_sc}
As discussed in the main text, we focus on the interacting tight-binding Hamiltonian of tWSe${}_2$ with non-interacting term $H_0$, displacement field term $H_D$, onsite Hubbard repulsion $H_U$, and NNN attraction within A and B sublattices $H_{V_2}$: 
\begin{equation}\label{HamiltonianV2}
    H = H_0 + H_D + H_U + H_{V_2}=\sum_{ij\alpha\beta\sigma} \left(t_{i\alpha j\beta\sigma}-\mu_\alpha\delta_{ij}\delta_{\alpha\beta}\right) c_{i\alpha\sigma}^\dagger c_{j\beta\sigma}  + \sum_{i\alpha} U_\alpha n_{i\alpha\uparrow} n_{i\alpha\downarrow} -V_2\sum_{i \sigma \sigma^\prime\alpha}\sum_{\in \{A,B\}} \sum_{\delta\in NNN} n_{i+\delta \alpha\sigma} n_{i\alpha\sigma^\prime},
\end{equation}
where we have combined the chemical potential and displacement field as $\mu_A = \mu - \mathcal{V}_z/2$, $\mu_B = \mu + \mathcal{V}_z/2$ and $\mu_C=\mu$ for simplicity, and other notations are the same as the main text.

The NNN attraction $H_{V_2}$ generally leads to a SC phase, while the onsite Hubbard repulsion $H_U$ typically favors magnetic ordered phase ($120^\circ$ AFM order in this case). Since the SC phase and magnetic ordered phase typically do not coexist, here we assume vanishing magnetic order in the SC mean-field analysis. Thus, according to Eq.~\eqref{mfhubbard} in the main text, the mean-field decoupling of $H_U$ only contributes an additional sublattice potential term $\frac{1}{2}\sum_{i\alpha}U_\alpha n_{i\alpha}$.

We decouple the NNN attraction $H_{V_2}$ in the SC channel as:
\begin{equation}\label{mfreal}
    \begin{aligned}
        H_{V_2} \approx -V_2\sum_{i \sigma \sigma^\prime} \sum_{\alpha\in \{A,B\}} \sum_{\delta\in NNN} \left( 
      \tilde{\Delta}_{\alpha\sigma\sigma^\prime}^* (\delta)c_{i\alpha\sigma^\prime} c_{i+\delta \alpha\sigma} 
    +  c_{i+\delta \alpha \sigma }^\dagger c_{i\alpha \sigma^\prime}^\dagger \tilde{\Delta}_{\alpha\sigma\sigma^\prime} (\delta)
    - \tilde{\Delta}_{\alpha\sigma\sigma^\prime}^* (\delta) \tilde{\Delta}_{\alpha\sigma\sigma^\prime}(\delta)
    \right),
    \end{aligned}
\end{equation}
where we have defined the spatially uniform real space pairing order parameter as $\tilde{\Delta}_{\alpha\sigma\sigma^\prime} (\delta)= \langle c_{i\alpha\sigma^\prime} c_{i+\delta\alpha\sigma} \rangle$. 

The dependence of order parameter $\tilde{\Delta}_{\alpha\sigma\sigma^\prime} (\delta)$ on NNN bond direction $\delta$ is classified according to the SC pairing symmetries. In the main text, we have classified the $S_z=0$ inter-valley pairing based on the irreducible representation of the point group $C_{3v}$. Additionally, here we also consider the case that the SC phase breaks $C_3$ rotation symmetry of the system (i.e. nematic SC) in E representation of $C_{3v}$, under the assumption that the pairing still preserves one of the three mirror symmetry of $C_{3v}$. Tab.~\ref{tab:pairsym} summarizes the $C_{3v}$ irreducible representations and the form factors $f(\delta)$ (defined by $\tilde{\Delta}_{\alpha\sigma\sigma^\prime} (\delta)=f(\delta)\tilde{\Delta}_{\alpha\sigma\sigma^\prime}$) for each pairing symmetries we considered. And in the following calculations, we will examine each pairing symmetries separately, comparing their energies to determine the most favorable one.
\begin{table}[h]
    \centering
    \begin{tabular}{c|c|c|c|c}
        \hline\hline
        Pairing Symmetry  & \makecell{Irreducible\\Representation}&$f(\delta_1)$ & $f(\delta_2)$ & $f(\delta_3)$ \\ \hline
        Mixed $s$-$f$ wave&$A_1$&$1$&$1$&$1$\\ \hline
        Mixed $p_x\pm ip_y$ - $d_{xy}\mp id_{x^2-y^2}$ wave&$E$&$1$&$e^{\pm i \frac{2\pi}{3}}$&$e^{\mp i \frac{2\pi}{3}}$\\ \hline
        Mixed $p_x$ - $d_{xy}$ wave&$E$&$0$&$\frac{\sqrt{3}}{2}$&$-\frac{\sqrt{3}}{2}$\\ \hline
        Mixed $p_y$ - $d_{x^2-y^2}$ wave&$E$&$1$&$-\frac{1}{2}$&$-\frac{1}{2}$\\ \hline\hline
    \end{tabular}
    \caption{Details of SC pairing symmetries on NNN bonds}
    \label{tab:pairsym}
\end{table}

With the above symmetry considerations, we take $\sigma^\prime=\bar{\sigma}$ to be opposite to $\sigma$ for $S_z=0$ inter-valley pairing, and then Fourier transform the real space mean-field decoupling Eq.~\eqref{mfreal} to momentum space:
\begin{equation}
    \begin{split}
        & -V_2\sum_{i \sigma } \sum_{\alpha\in \{A,B\}} \sum_{\delta\in NNN} \left( 
      \tilde{\Delta}_{\alpha\sigma\bar{\sigma}}^*(\delta) c_{i\alpha\bar{\sigma}} c_{i+\delta \alpha\sigma} 
    +  c_{i+\delta \alpha \sigma }^\dagger c_{i\alpha \bar{\sigma}}^\dagger \tilde{\Delta}_{\alpha\sigma\bar{\sigma}}(\delta)
    - \tilde{\Delta}_{\alpha\sigma\bar{\sigma}}^*(\delta) \tilde{\Delta}_{\alpha\sigma\bar{\sigma}}(\delta)
    \right) \\
    = & \sum_{k \sigma } \sum_{\alpha\in \{A,B\}} \left(
    \Delta_{\alpha\sigma\bar{\sigma}}^*(k) c_{-k\alpha\bar{\sigma}} c_{k \alpha\sigma} 
    + c_{k \alpha \sigma }^\dagger c_{-k\alpha \bar{\sigma}}^\dagger \Delta_{\alpha\sigma\bar{\sigma}}(k) 
    \right) 
    + NV_2 \sum_{\sigma } \sum_{\alpha\in \{A,B\}} \sum_{\delta\in NNN} \tilde{\Delta}_{\alpha\sigma\bar{\sigma}}^*(\delta) \tilde{\Delta}_{\alpha\sigma\bar{\sigma}}(\delta)
    \end{split}
\end{equation}
where we have defined the momentum space gap function $\Delta_{\alpha\sigma\bar{\sigma}}(k)$ as $\Delta_{\alpha\sigma\bar{\sigma}}(k) \equiv -V_2 \tilde{\Delta}_{\alpha\sigma\bar{\sigma}} \sum_{\delta\in NNN} f(\delta) e^{-ik\delta}$, and the self-consistency relation in momentum space is given by:
\begin{equation}\label{selfconsistmomentum}
    \tilde{\Delta}_{\alpha\sigma\bar{\sigma}}(\delta) = \frac{1}{N} \sum_k \langle c_{-k\alpha\bar{\sigma}} c_{k\alpha \sigma} \rangle e^{ik\delta}.
\end{equation}

Together with the Fourier transformation of $H_0 + H_D$, we can then write the full momentum space mean-field Hamiltonian into the Bogoliubov-de-Gennes(BdG) form:
\begin{equation}\label{bdg}
\begin{aligned}
    H_\text{MF} =&  \sum_{k\alpha\beta\sigma} \left(\epsilon_{\alpha\beta\sigma}(k) -\mu _\alpha\delta_{\alpha\beta}\right)c_{k\alpha\sigma}^\dagger c_{k\beta\sigma}  +
    \sum_{k \sigma } \sum_{\alpha\in \{A,B\}} \left(
    \Delta_{\alpha\sigma\bar{\sigma}}^*(k) c_{-k\alpha\bar{\sigma}} c_{k \alpha\sigma} 
    + c_{k \alpha \sigma }^\dagger c_{-k\alpha \bar{\sigma}}^\dagger \Delta_{\alpha\sigma\bar{\sigma}}(k) 
    \right) \\
    &+ NV_2 \sum_{\sigma } \sum_{\alpha\in \{A,B\}} \sum_{\delta\in NNN} \tilde{\Delta}_{\alpha\sigma\bar{\sigma}}^*(\delta) \tilde{\Delta}_{\alpha\sigma\bar{\sigma}}(\delta) \\
    =& \frac{1}{2}\sum_{k}
    \begin{pmatrix}
        \bm{c}_{k\uparrow}^\dagger & \bm{c}_{k\downarrow}^\dagger & \bm{c}_{-k\uparrow} & \bm{c}_{-k\downarrow}
    \end{pmatrix}
    \begin{pmatrix}
        \bm{\epsilon}_\uparrow(k) - \bm{\mu} & \bm{0} & \bm{0} & \bm{\Delta}_{\uparrow\downarrow}(k)\\
        \bm{0} & \bm{\epsilon}_\downarrow(k) - \bm{\mu} & \bm{\Delta}_{\downarrow\uparrow}(k) & \bm{0}\\
        \bm{0} & \bm{\Delta}_{\downarrow\uparrow}^\dagger(k) & -\bm{\epsilon}^T_\uparrow(-k)+\bm{\mu} & \bm{0}\\
        \bm{\Delta}_{\uparrow\downarrow}^\dagger(k) & \bm{0} & \bm{0} &-\bm{\epsilon}_\downarrow^T(-k)+\bm{\mu}
    \end{pmatrix}
     \begin{pmatrix}
        \bm{c}_{k\uparrow} \\ \bm{c}_{k\downarrow} \\ \bm{c}_{-k\uparrow}^\dagger \\ \bm{c}_{-k\downarrow}^\dagger
    \end{pmatrix} \\
    &+ \frac{1}{2} \sum_{k\alpha\sigma} \left(\epsilon_{\alpha\alpha\sigma}(k)-\mu_\alpha  \right)
    + NV_2 \sum_{\sigma } \sum_{\alpha\in \{A,B\}} \sum_{\delta\in NNN} \tilde{\Delta}_{\alpha\sigma\bar{\sigma}}^*(\delta) \tilde{\Delta}_{\alpha\sigma\bar{\sigma}}(\delta),
\end{aligned}
\end{equation}
where the pairing matrix is defined as:
\begin{equation}
    \bm{\Delta}_{\sigma\bar{\sigma}}(k) = 
    \begin{pmatrix}
        \Delta_{A\sigma\bar{\sigma}}(k) - \Delta_{A\bar{\sigma}\sigma}(-k) &0&0\\
        0&\Delta_{B\sigma\bar{\sigma}}(k) - \Delta_{B\bar{\sigma}\sigma}(-k)&0\\
        0&0&0
    \end{pmatrix}
\end{equation}

With the mean-field Hamiltonian derived above, we can perform the standard mean-field calculations. We first construct the BdG Hamiltonian in Eq.~\eqref{bdg} using some randomly chosen initial values of the order parameters $\tilde{\Delta}_{\alpha\sigma\bar{\sigma}}$, and adjust the chemical potential $\mu$ to fit the correct average filling $\nu=-1$ by $n_e = \sum_{k\alpha\sigma}\langle c_{k\alpha\sigma}^\dagger  c_{k\alpha\sigma}\rangle$, with the many-body ground state of the BdG Hamiltonian is obtained by filling all quasi-particle states with negative energy. And the ground state expectation value in Eq.~\eqref{selfconsistmomentum} is calculated to update the order parameters for the next iteration.  The above process is repeated until the self-consistency is achieved.

Although we have argued in the main text that NNN attraction is more likely to induce stronger SC instability, here we also examine an alternative scenario that the NNN attraction $H_{V_2}$ is replaced by a NN attraction $H_{V_1}$:
\begin{equation}\label{HamiltonianV1}
    H^\prime = H_0 + H_D + H_U + H_{V_1}=\sum_{ij\alpha\beta\sigma} \left(t_{i\alpha j\beta\sigma}-\mu_\alpha\delta_{ij}\delta_{\alpha\beta}\right) c_{i\alpha\sigma}^\dagger c_{j\beta\sigma}  + \sum_{i\alpha} U_\alpha n_{i\alpha\uparrow} n_{i\alpha\downarrow} -V_1\sum_{i \sigma \sigma^\prime} \sum_{\delta\in NN} n_{i+\delta A\sigma} n_{iB\sigma^\prime}. 
\end{equation}

The SC mean-field analysis for $H^\prime$ closely mirrors that of $H$ we have derived above, where the only differences are the substitution of $\delta$ from NNN to NN bonds and the modification of sublattice indices. Therefore, we will omit the derivation here for brevity, and only show the corresponding classification of SC pairing symmetries on NN bonds (only considering chiral SC in E representation) in Tab.~\ref{tab:pairsymnn}.
\begin{table}[h]
    \centering
    \begin{tabular}{c|c|c|c|c}
        \hline\hline
        Pairing Symmetry  & \makecell{Irreducible\\Representation}&$f(\delta_1)$ & $f(\delta_2)$ & $f(\delta_3)$ \\ \hline
        $s$ wave&$A_1$&$1$&$1$&$1$\\ \hline
        Mixed $p_x\pm ip_y$ - $d_{xy}\mp id_{x^2-y^2}$ wave&$E$&$1$&$e^{\pm i \frac{2\pi}{3}}$&$e^{\mp i \frac{2\pi}{3}}$\\ \hline\hline
    \end{tabular}
    \caption{Details of SC pairing symmetries on NN bonds}
    \label{tab:pairsymnn}
\end{table}

\begin{figure}[h] 
\centering 
\includegraphics[width=0.95\textwidth]{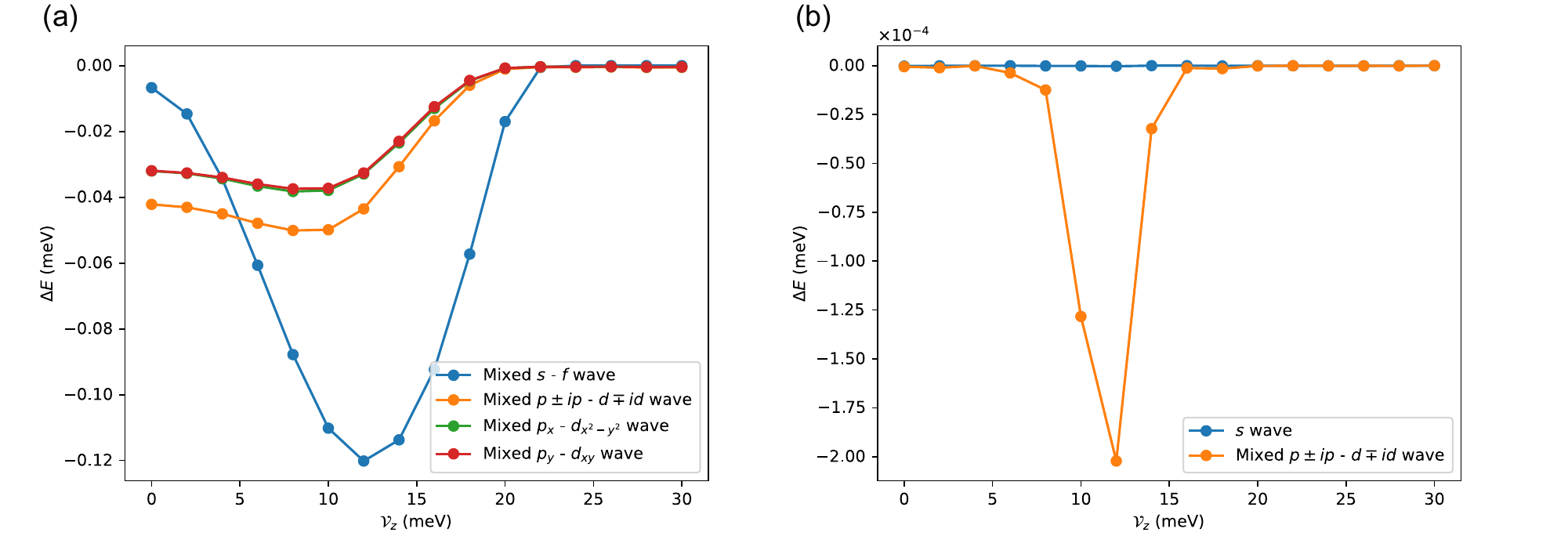} 
\caption{The energy gain $\Delta E$ per hole of the SC phases compared to the symmetric phase with (a) NNN and (b) NN pairings.}
\label{fig:energy_gain_supp}
\end{figure}

Taking $U_A = U_B =U_C = 37.5$ meV, $V_2 = 10$ meV as in the main text for the NNN Hamiltonian $H$ (see Eq.~\eqref{HamiltonianV2}), and $V_1 = 10$ meV for the NN Hamiltonian $H^\prime$ (see Eq.~\eqref{HamiltonianV1}), we perform the mean-field calculations for all pairing symmetries classified in Tab.~\ref{tab:pairsym} and Tab.~\ref{tab:pairsymnn}. The resulting energy gain per hole for NNN and NN Hamiltonians are illustrated in Fig.~\ref{fig:energy_gain_supp}(a) and (b) respectively. 

A direct comparison of the energy gains $\Delta E$ reveals that NNN pairings are approximately $2\sim 3$ orders of magnitude stronger than NN pairings for same attraction strength $V_1=V_2$. We thus conclude that the NNN pairings should be more dominant than the NN pairings, justifying our main focus of the NNN pairings in the main text. This result is actually straightforward to understand. Due to the time-reversal symmetry $\mathcal{T}$ in both $H$ and $H^\prime$, the Cooper pairs will form between time-reversal-related states. As the time-reversal symmetry $\mathcal{T}$ does not alter sublattice index $\alpha$, and given that the Fermi surface have relatively strong sublattice polarization property (as shown in Fig.~\ref{fig:fermisurface}(a) in the main text), the system is expected to favor intra-sublattice pairings (e.g. NNN pairings) over inter-sublattice pairings (e.g. NN pairings). Considering the layer polarization property of the A and B sublattices (the main components of the Fermi surface), it is equivalent to say that the tWSe${}_2$ system favors intra-layer pairings. 

\begin{figure}[h] 
\centering 
\includegraphics[width=0.95\textwidth]{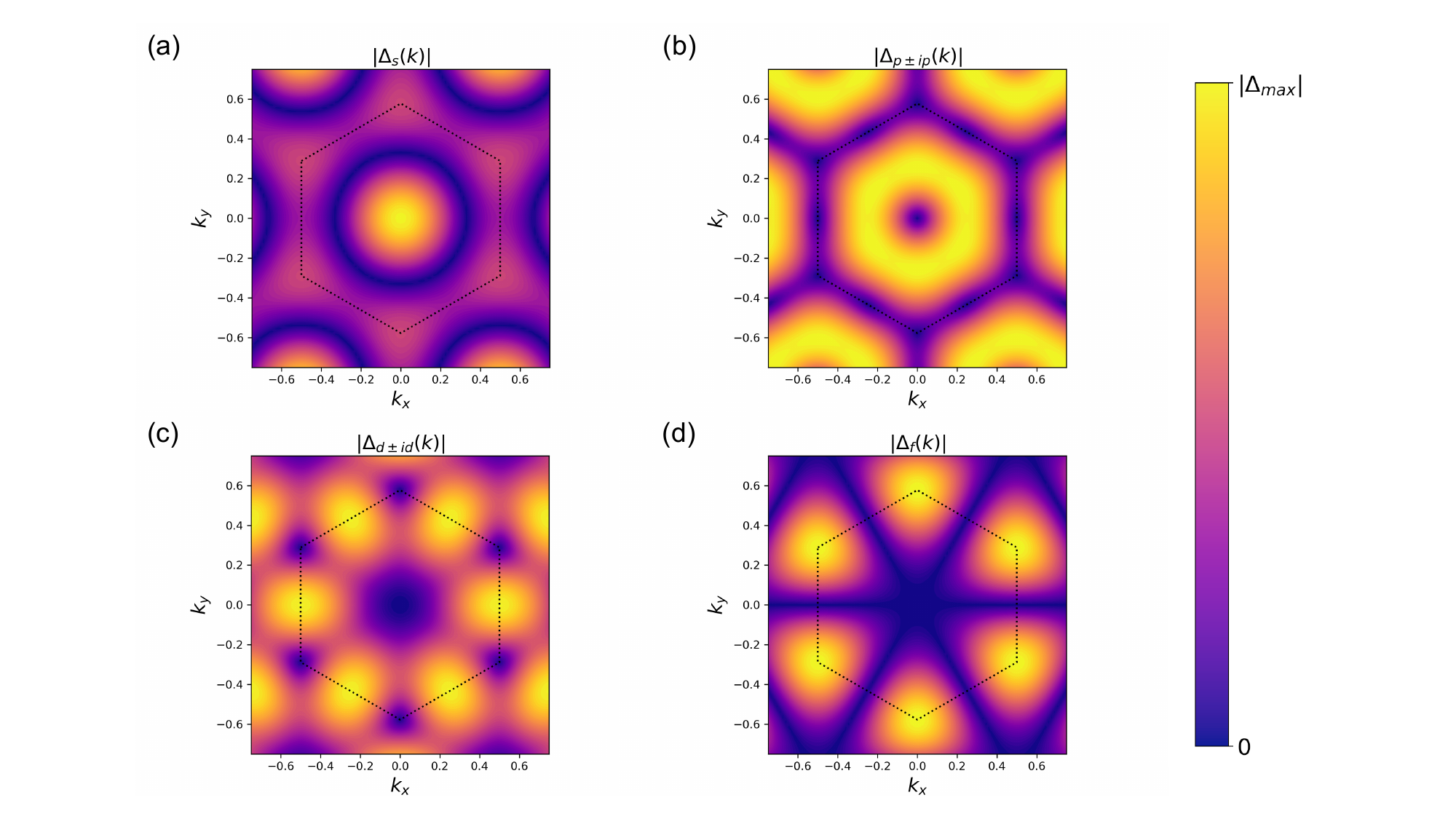} 
\caption{Momentum space form factor for NNN pairing with (a) s-wave, (b) $p_x\pm ip_y$-wave, (c) $d_{xy}\mp id_{x^2-y^2}$-wave and (d) $f$-wave symmetry, and the BZ is shown as black dash line in each figure.}
\label{fig:form}
\end{figure}

Next, we examine the nematic SC orders (with mixed $p_x$-$d_{xy}$ or $p_y$-$d_{x^2-y^2}$-wave symmetry) as shown in Fig.~\ref{fig:energy_gain_supp}(a). These nematic SC states have energies considerably higher than the chiral ones under small displacement field. It is expected that some moderate $C_3$ breaking terms (e.g. strain) may favor such nematic SC over chiral ones. And it should be also noted that these two nematic SC states are nearly (but not exactly) degenerate in energy.

Returning to the non-nematic cases for NNN pairings, one can see that at small $\mathcal{V}_z$ (as discussed in the main text) or large enough $\mathcal{V}_z$ regime (where the energies of different pairing symmetries are close), the chiral mixed $p_x\pm ip_y$ - $d_{xy}\mp id_{x^2-y^2}$-wave symmetry is the most dominant, while at intermediate $\mathcal{V}_z$, the mixed $s$-$f$-wave symmetry becomes the strongest one. To understand this, we plot the magnitude of the momentum space form factor for NNN pairings with different partial wave components in Fig.~\ref{fig:form}. As shown in Fig.~\ref{fig:form}, the $s$-wave component has local maximum at $\gamma$ and $\kappa_\pm$ points of Brillouin zone (BZ) while having ring-shaped minimum in between; the $p_x\pm ip_y$-wave component is large except at $\gamma$ points and the BZ boundary; the $d_{xy}\pm id_{x^2-y^2}$-wave component has maximum at $m$ points while having minimum at $\gamma$ and $\kappa_\pm$ points; and the $f$ wave component has maximum at $\kappa_\pm$ points but having minimum nodal line along $\gamma$-$m$ directions.

The energetic of these different SCs can then be understood by combining the Fermi surface structures shown in Fig.~\ref{fig:fermisurface} in the main text with the form factors shown in Fig.~\ref{fig:form}. At small displacement field $\mathcal{V}_z$, the Fermi surface is hexagonal shape around the $\gamma$ point in the BZ, where the chiral $p_x\pm ip_y$ and $d_{xy}\pm id_{x^2-y^2}$ components are the largest and fully gaps the Fermi surface. When increasing $\mathcal{V}_z$, Fig.~\ref{fig:fermisurface}(b) in the main text indicates that B sublattice has dominant DOS, which should be our primary focus. And Fig.~\ref{fig:fermisurface}(a) in the main text clearly illustrates that the B pockets are move towards $\kappa_\pm$ points in the BZ as the displacement field $\mathcal{V}_z$ is increasing, where the $s$ and $f$-wave components are larger in magnitude and become better in energy. However, as $\mathcal{V}_z$ increases further, the B sublattice Fermi pockets at $\kappa_\pm$ points disappear, leaving the remaining Fermi surface mainly on A sublattice, which is again located at the region where $p_x\pm ip_y$ and $d_{xy}\pm id_{x^2-y^2}$ components are dominant. The above analysis suggests that the combined use of the structure of Fermi surfaces and momentum space pairing form factors provides a clear explanation of the mean-field behaviors of the SC phases.

\section{Derivations of Magnetic Mean-Field Analysis}\label{supp:mf_afm}
In this section, we provide a more detailed mean-field analysis for possible magnetic orders of the NNN Hamiltonian $H$ in Eq.~\eqref{HamiltonianV2}. Assuming the absence of SC order (hence $H_{V_2}$ has vanishing mean-field contributions), we mainly focus on the in-plane magnetic channel of $H_U$ as:
\begin{equation}\label{equ:mfhubbard}
    H_U \approx -\sum_{i\alpha} U_\alpha \left(
       m_{i\alpha} c_{i\alpha\downarrow}^\dagger c_{i\alpha\uparrow} +  m_{i\alpha}^* c_{i\alpha\uparrow}^\dagger c_{i\alpha\downarrow} \right)+\sum_{i\alpha}U_\alpha |m_{i\alpha}|^2 + \frac{1}{2} \sum_{i\alpha\sigma}U_\alpha n_{i\alpha\sigma},
\end{equation}
where $m_{i\alpha} = \langle c_{i\alpha\uparrow}^\dagger c_{i\alpha\downarrow} \rangle = \langle S_{i\alpha}^x\rangle + i\langle S_{i\alpha}^y\rangle$ is the in-plane magnetic order parameter. As discussed in the main text, due to the existence of approximate nesting wave vector $\bm{Q}=(0, 4\pi/3a_M)$ between spin up and down Fermi surfaces, we first consider the ansatz of $120^{\circ}$ AFM order $m_{i\alpha} = m_{\alpha}^+ e^{iQ\cdot R_i} + m_{\alpha}^- e^{-iQ\cdot R_i}=\sum_{\eta}m_{\alpha}^\eta e^{i\eta Q\cdot R_i}$. Performing Fourier transformation, the mean-field Hamiltonian is given by:

\begin{equation}\label{afmmfhamiltonian}
H_\text{MF} = \sum_{k\alpha\beta\sigma} \left(\epsilon_{\alpha\beta\sigma}(k) + \frac{U_\alpha}{2}\delta_{\alpha\beta} -\mu_\alpha\delta_{\alpha\beta}\right)c_{k\alpha\sigma}^\dagger c_{k\beta\sigma}   - \sum_{k\alpha\eta} U_\alpha \left( m_\alpha^\eta  c_{k+\eta Q\alpha\downarrow}^\dagger c_{k\alpha\uparrow} 
        + m_\alpha^{\eta*} c_{k-\eta Q\alpha\uparrow}^\dagger c_{k\alpha\downarrow} 
        \right)
        +N\sum_{\alpha\eta}U_\alpha |m_{\alpha}^{\eta}|^2 ,
\end{equation}
and the order parameter $m_{\alpha}^\eta$ satisfies the self-consistent equation: 
\begin{equation}\label{equ:consistence_afm}
    m_{\alpha}^\eta=\frac{1}{N}\sum_{k}\langle c_{k\alpha\uparrow}^{\dagger}c_{k+\eta Q\alpha\downarrow}\rangle.
\end{equation}

Notice that the momentum summation above should be performed in the original BZ with $N$ being the number of original unit cells. And in practice, the folding of BZ, as shown in Fig.~\ref{fig:afm}(d) in the main text, can be easily achieved by a simple relabeling of momentum.

Then, we can perform the standard mean-field calculations as follow. We first construct and diagonalize the mean-field Hamiltonian in Eq.~\eqref{afmmfhamiltonian}, and filled the lowest energy states until $\nu=-1$. We then use Eq.~\eqref{equ:consistence_afm} to update the order parameters for next iteration. Such process is repeated until self-consistency is achieved, and the numerical results are detailed in the main text.

However, there are also other possible competing orders, one important class being the zero momentum magnetic orders, which do not break the translation symmetry. Since there is no experimental evidence of valley polarization ferromagnetism (i.e. magnetic order in $z$ direction), here we again focus on the in-plane magnetic channel. The decoupling of the Hubbard interaction $H_U$ in momentum space is:

\begin{equation}
    H_{U}\approx - \sum_{k\alpha} U_{\alpha} \left(m_{\alpha} c_{\alpha k \downarrow}^{\dagger} c_{\alpha k \uparrow}+m_{\alpha}^*c_{\alpha k \uparrow}^{\dagger} c_{\alpha k \downarrow}\right)+N\sum_{\alpha}U_\alpha m_{\alpha}^2+  \frac{1}{2} \sum_{k\alpha\sigma}U_\alpha n_{k\alpha\sigma},
\end{equation}
where the in-plane magnetic order parameter is given by $m_{\alpha}=\frac{1}{N}\sum_{k}\langle c_{k\alpha\uparrow}^{\dagger}c_{k\alpha\downarrow}\rangle$. We have numerically checked that such magnetic order is not favored when turning the displacement field on, and we conclude that $120^{\circ}$ AFM order is indeed more favorable. Such results can be easily understood since the displacement field splits the Fermi surfaces of spin up and spin down, hence suppress the zero-momentum inter-valley coherent magnetic orders.

\section{Mean-Field Results Away From Filling $\nu=-1$}\label{supp:mf_filling}

In the experiment, the $3.65^\circ$ tWSe$_2$ system exhibits metallic behavior on both sides as the filling deviating from $\nu=-1$. In this section, we present a preliminary theoretical investigation on the doping dependence of tWSe$_2$ near the SC phase through mean-field analysis, demonstrating that our model can also explain the key experimental features when the filling is away from $\nu=-1$.
Theoretically, the commensurate $\sqrt{3} \times \sqrt{3}$ AFM order tendency generally persists over a certain range of filling near $\nu=-1$, leading to a metallic phase since the filling factor (in the enlarged unit cell) deviates from the integer value $\tilde{\nu}=-3$. Meanwhile, the SC tendency also remains robust upon doping away from $\nu=-1$, as the time-reversal symmetry remains intact. The actual ground state is determined by the energetics of these two phases.

\begin{figure}[t] 
\centering 
\includegraphics[width=1.0\textwidth]{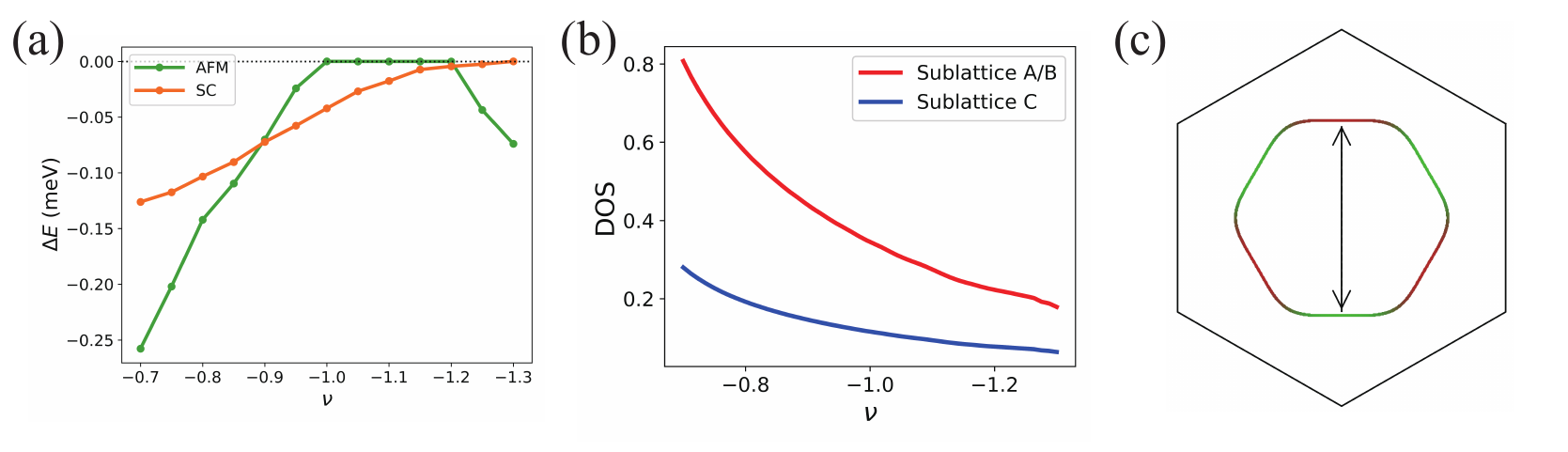} 
\caption{(a) The energy gain per hole of the AFM order and the mixed $p_x\pm ip_y$ and $d_{xy}\mp id_{x^2-y^2}$-wave SC order. (b) the Fermi surface DOS as a function of filling $\nu$. (c) Fermi surface at $\nu=-1.3$, where the nesting wave vector $\bm{Q}=(0, \pm4\pi/3a_M)$ is represented as black arrow.}
\label{fig:changing_f}
\end{figure}

Fig.~\ref{fig:changing_f}(a) illustrates our mean-field results for the filling dependence of SC and AFM orders at $\mathcal{V}_z = 0$ near $\nu=-1$, indicating that the SC phase persists only within the range $0.9 \lesssim |\nu| \lesssim 1.2$, while an AFM metal phase emerges beyond this filling range, qualitatively consistent with the experimental observations. To better understand the filling dependence of these orders, we present the Fermi surface DOS as a function of $\nu$ in Fig.~\ref{fig:changing_f}(b), where the DOS is significantly enhanced (suppressed) on the less (more) hole-doped side due to its closer (further) proximity to the van Hove singularity. The Fermi surface DOS provides a natural understanding for the strength of SC order, as well as the enhanced AFM order on the less hole-doped side. In contrast, the emergence of AFM order on the more hole-doped side, as shown in Fig.~\ref{fig:changing_f}(c), arises from the better nesting condition for the $\sqrt{3} \times \sqrt{3}$ AFM order.

\end{document}